\documentclass[reprint,prb,showkeys,superscriptaddress,citeautoscript] {revtex4-2}

\usepackage{graphicx}
\usepackage{tikz}
\usepackage{braket}
\usepackage{hyperref}
\hypersetup{colorlinks=true, urlcolor= blue, citecolor=blue, linkcolor= blue, bookmarks=true, bookmarksopen=false}

\usepackage{amsmath,amssymb}
\usepackage{textcomp}
\usepackage{multirow}
\usepackage{enumitem}
\setlist{noitemsep, leftmargin=*}
\usepackage{times}
\begin{document}
		
		\title{ Mechanical, optical, and thermoelectric properties of  semiconducting ZnIn$_2$X$_4$ (X= S, Se, Te) monolayers}
		
		\author{Mohammad Ali Mohebpour}
		\email{b92\_mohebi@hotmail.com}
		\affiliation{Department of Physics, University of Guilan, P. O. Box 41335-1914, Rasht, Iran}
		
		\author{Bohayra Mortazavi}  
		\email{bohayra.mortazavi@gmail.com}
		\affiliation{Department of Mathematics and Physics, Leibniz Universit{\"a}t Hannover, Appelstra$\beta$e 11,30167 Hannover, Germany}
		
		\author{Timon Rabczuk}
		\affiliation{College of Civil Engineering, Department of Geotechnical Engineering, Tongji University, 1239 Siping Road Shanghai, China}
		
		\author{Xiaoying Zhuang}
		\affiliation{Department of Mathematics and Physics, Leibniz Universit{\"a}t Hannover, Appelstra$\beta$e 11,30167 Hannover, Germany}
		\affiliation{College of Civil Engineering, Department of Geotechnical Engineering, Tongji University, 1239 Siping Road Shanghai, China}
		
		\author{Alexander V. Shapeev}
		\affiliation{Skolkovo Institute of Science and Technology, Skolkovo Innovation Center, Bolshoy Bulvar, 30s1, Moscow, 143026, Russia}
		
		\author{Meysam Bagheri Tagani}
		\email{m\_bagheri@guilan.ac.ir}
		\affiliation{Department of Physics, University of Guilan, P. O. Box 41335-1914, Rasht, Iran}

%%%%%%%%%%%%%%%%%%%%%%%%%%%%%%%%%%%%%%%%%%%%%%%%%%%%%%%%%%%%%%%%%%%%%
\begin{abstract}
		
In the latest experimental success in the field of two-dimensional materials, ZnIn$_2$S$_4$ nanosheets with a highly appealing efficiency for photocatalytic hydrogen evolution were synthesized (ACS Nano 15(2021), 15238). Motivated by this accomplishment, herein, we conduct first-principles-based calculations to explore the physical properties of the ZnIn$_2$X$_4$ (X= S, Se, Te) monolayers. The results confirm the desirable dynamical and mechanical stability of the ZnIn$_2$X$_4$ monolayers. The ZnIn$_2$S$_4$ and ZnIn$_2$Se$_4$ are semiconductors with direct band gaps of 3.94 and 2.77 eV, respectively whereas the ZnIn$_2$Te$_4$ shows an indirect band gap of \mbox{1.84 eV} at the G$_0$W$_0$ level. The optical properties achieved from the solution of the \mbox{Bethe-Salpeter} equation predict the exciton binding energy of the ZnIn$_2$S$_4$, ZnIn$_2$Se$_4$, and ZnIn$_2$Te$_4$ monolayers to be 0.51, 0.41, and \mbox{0.34 eV}, respectively, suggesting the high stability of the excitonic states against thermal dissociation. Using the iterative solutions of the Boltzmann transport equation accelerated by machine learning interatomic potentials, the \mbox{room-temperature} lattice thermal conductivity of the ZnIn$_2$S$_4$, ZnIn$_2$Se$_4$, and ZnIn$_2$Te$_4$ monolayers is predicted to be remarkably low as 5.8, 2.0, and 0.4 W/mK, respectively. Due to the low lattice thermal conductivity, high thermopower, and large figure of merit, we propose the ZnIn$_2$Se$_4$ and ZnIn$_2$Te$_4$ monolayers as promising candidates for thermoelectric energy conversion systems. This study provides an extensive vision concerning the intrinsic physical properties of the ZnIn$_2$X$_4$ nanosheets and highlights their characteristics for energy conversion and optoelectronics applications.

\end{abstract}
	
\maketitle

%%%%%%%%%%%%%%%%%%%%%%%%%%%%%%%%%%%%%%%%%%%%%%%%%%%%%%%%%%%%%%%%%%%%%
	
\section{Introduction}

Graphene \cite{novoselov2004electric, novoselov2007rise, neto2009electronic}, the full-sp$^2$ carbon atoms arranged in a planar honeycomb lattice, exhibits exceptional mechanical strength \cite{lee2008measurement}, high carrier mobility \cite{banszerus2015ultrahigh}, high thermal conductivity \cite{ghosh2008extremely, balandin2008superior}, and excellent electronic features \cite{berger2004ultrathin, liu2011graphene, withers2010electron, liu2019recent}. Graphene's successes promoted the field of two-dimensional (2D) materials which has been continuously expanding during the last decade. It is worth reminding that semiconducting materials with suitable electronic band gap are required for advanced mainstream technologies, such as optoelectronics, sensors, electronics, catalysis, and energy converters. Pristine graphene nonetheless shows an isotropic Dirac cone and zero electronic band gap, limiting its effectiveness for numerous cutting-edge technologies. The flexible nature of carbon atoms allows the structure of graphene to be manipulated so that a semiconductor, as graphdiyne \cite{baughman1987structure} or an insulator as fluorinated graphene \cite{bakharev2020chemically} was synthesized. Graphene physics also offers the possibility of the band gap opening by chemical functionalization \cite{wehling2008molecular, schedin2007detection, soriano2015spin}, defect engineering \cite{lherbier2012transport}, or mechanical straining \cite{si2016strain, guinea2012strain}.

For the cost-effective practical applications, it is more appealing to employ 2D intrinsic semiconductors rather than the band gap opening in graphene. That is why, many 2D semiconductors have been fabricated up to now, such as: single-triazine-based g-C$_3$N$_4$ \cite{algara2014triazine}, transition metal dichalcogenides family \cite{geim2013van, wang2012electronics, radisavljevic2011single}, phosphorene \cite{das2014ambipolar, li2014black}, indium selenide \cite{bandurin2017high}, MoSi$_2$N$_4$ family \cite{hong2020chemical}, polyaniline C$_3$N \cite{mahmood2016two}, graphene-like BC$_2$N \cite{seo2021dominant}, nickel diazenide NiN$_2$ \cite{bykov2021realization}, niobium oxide diiodide NbOI$_2$ \cite{fang20212d}, and most recently penta-PdPS \cite{wang2022polarimetric} and PdPSe \cite{li2021penta} nanosheets. Owing to the widespread applications of 2D semiconductors in different technologies, tremendous experimental endeavors are continuously devoted to designing and synthesizing novel nanosheets with improved performances. For example, to enhance the efficiency in the thermoelectric energy conversion, the utilized semiconductors ought to simultaneously show low lattice thermal conductivity and high electrical conductivity.

In the continued effort of the prediction and experimental fabrication of 2D semiconductors, most recently, Zhang \mbox{et al.} \cite{Zhang2021ACS} succeeded in fabricating the layered structure of ZnIn$_2$S$_4$ using a hydrothermal method. They found that the ZnIn$_2$S$_4$ 2D system can be employed for photocatalytic hydrogen evolution. This latest advance is also expected to facilitate the synthesis of the ZnIn$_2$Se$_4$ and ZnIn$_2$Te$_4$ nanosheets with similar atomic structures.
In this paper, we examine the stability and intrinsic physical properties of the ZnIn$_2$X$_4$ \mbox{(X= S, Se, Te)} monolayers. For this purpose, we perform the density functional theory calculations to investigate the mechanical, optoelectronic, and thermoelectric properties. The excitonic optical properties of the monolayers are calculated by solving the Bethe-Salpeter equation. The lattice thermal conductivity are predicted by employing the full-iterative solutions of the Boltzmann transport equation, accelerated by machine learning interatomic potentials. The electronic transport properties are calculated by the semiclassical Boltzmann transport equation within the relaxation time approximation. The obtained results reveal a decrease in the elastic modulus, tensile strength, phonon group velocity, phonon lifetime, lattice thermal conductivity, and exciton binding energy with the increase in the atomic weight of chalcogen atom in the ZnIn$_2$X$_4$ nanosheets. This work provides a comprehensive vision on the stability and key physical properties of the ZnIn$_2$X$_4$ monolayers and highlights their prospect to design next-generation optoelectronic and energy conversion nanodevices.

\section {Computational methods}

The first-principles density functional theory (DFT) calculations are performed by employing the Vienna \mbox{Ab-initio} Simulation Package \cite{kresse1996efficient, perdew1996generalized}. The generalized gradient approximation (GGA) is employed with the Perdew-Burke-Ernzerhof (PBE) exchange-correlation functional. The plane wave and self-consistent loop cutoff energies are defined as 500 and $10^{-6}$ eV, respectively. To optimize the structures, the atomic positions and lattice sizes are fully relaxed using the conjugate gradient algorithm until the Hellman-Feynman forces drop below $10^{-3}$ eV/\AA. The Brillouin zone (BZ) is integrated with a 7$\times$7$\times$1 Monkhorst-Pack \cite{monkhorst1976special} K-point grid. For the bulk structures, the DFT-D3 \cite{grimme2010consistent} dispersion correction by Grimme is adopted to account for the van der Waals interactions. The periodic boundary conditions are considered in all the directions for all  structures. For the monolayers, around 15 \AA~vacuum distance is introduced along the thickness to avoid interactions with systems' periodic images. The electronic structure is also analyzed by employing the HSE06 hybrid functional \cite{krukau2006influence}.

The electronic transport coefficients are calculated by solving the semiclassical Boltzmann transport equation (SBTE) within the relaxation time approximation (RTM) \cite{scheidee} as implemented in the BoltzTraP code \cite{madsen2006bol}. The electrical conductivity ($\sigma$) and the thermopower ($S$) are calculated as follows:
\begin{equation}
\label{sigma}
\sigma (\mu, T)=e^2 \sum_{\textbf{k}} {\tau_\textbf{k}} \int d\epsilon (-\frac{\partial f_\mu (\epsilon, T)}{\partial \epsilon}) v_{\textbf{k}} (\epsilon) \times v_{\textbf{k}} (\epsilon),
\end{equation}
	
\begin{equation}
\label{thermopower}
S (\mu, T)=\frac{ek_B}{\sigma} \sum_{\textbf{k}} {\tau_\textbf{k}} \int d\epsilon (-\frac{\partial f_\mu (\epsilon, T)}{\partial \epsilon}) v_{\textbf{k}} (\epsilon) \times v_{\textbf{k}} (\epsilon) \frac{\epsilon-\mu}{k_BT}, 
\end{equation}
where $f_\mu$ denotes the Fermi-Dirac distribution function.

The many-body perturbation theory (MBPT) calculations are performed using the non-self-consistent version of the GW method, referred to as single-shot G$_0$W$_0$. Herein, the quasiparticle (QP) band gap is attentively converged with respect to the number of conduction bands, number of frequency grid points, Brillouin zone mesh, and cutoff energy for the plane wave and the response function. After the convergence test, it was found that at least 192, 272, 352 virtual bands are essential for the ZnIn$_2$S$_4$, ZnIn$_2$Se$_4$, and ZnIn$_2$Te$_4$ monolayers, respectively to reach the convergence threshold of \mbox{10$^{-3}$ eV}. In addition, the number of frequency grid points is set to be 96. The Brillouin zone is integrated with a relatively denser \mbox{K-point} mesh of 9$\times$9$\times$1 to converge the G$_0$W$_0$ band gap within \mbox{10$^{-2}$ eV}. The cutoff energy for the plane wave and the response function are considered to be 500 and \mbox{200 eV}, respectively. The excitonic optical properties are investigated by calculating the frequency-dependent dielectric function, given as \mbox{$\epsilon(\omega)=\epsilon_1(\omega)+i\epsilon_2(\omega)$}, through solving the Bethe-Salpeter equation (BSE) over the G$_0$W$_0$ eigenvalues (the so-called G$_0$W$_0+$BSE). In this regard, the Tamm-Dancoff approximation (TDA) in used to simplify the BSE Hamiltonian by excluding the resonant-antiresonant coupling. The 15 highest valence bands and the 15 lowest conduction bands are included in the BSE calculations to achieve a converged spectrum. To show the effects of many-body interactions, the optical coefficients are also computed at lower levels of theory using the random-phase approximation (RPA) over the eigenvalues of the DFT
\mbox{(the so-called DFT$+$RPA)} and the G$_0$W$_0$
\mbox{(the so-called G$_0$W$_0+$RPA)}.

The density functional perturbation theory (DFPT) calculations are employed to obtain the phonon dispersions and harmonic force constants using the PHONOPY code \cite{togo2015first}. The moment tensor potentials (MTPs) \cite{shapeev2016moment} as an accurate class of machine learning interatomic potentials are used to interpolate the interatomic forces \cite{mortazavi2020exploring} utilizing the MLIP package \cite{novikov2020mlip}. The datasets for the MTPs training are achieved by conducting ab-initio molecular dynamics (AIMD) simulations with the time step of 1 fs over supercells consisting of 84 atoms using a 2$\times$2$\times$1 Monkhorst-Pack \mbox{K-point} grid. For evaluating the 2$^{nd}$ and 3$^{rd}$ order interatomic force constants, two AIMD calculations are performed within the NVT ensemble, first, from 10 to 100 K and second, from 100 to 1000 K, each for 1000 time steps. For the efficient training of the MTPs, the original AIMD trajectories are subsampled with equal steps and around 690 configurations are selected to train MTPs. Phonon dispersions on the basis of the trained MTPs are obtained using the PHONOPY code, as elaborately discussed in our previous work \cite{mortazavi2020exploring}. Anharmonic 3$^{rd}$ order interatomic force constants are obtained over the same supercells as those employed for harmonic force constant calculations by considering the interactions with the eighth nearest neighbors. The ShengBTE \cite{li2014shengbte} package is employed to perform the full iterative solution of the Boltzmann transport equation (BTE) with force constant inputs, as discussed in our previous study \cite{mortazavi2021accelerating}. We consider isotope scattering to predict the phononic thermal conductivity of the  samples.

\section{Results and Discussion}
\subsection{Structural and electronic properties}

\begin{figure*}
	\centering{
		\includegraphics[width=1\textwidth]{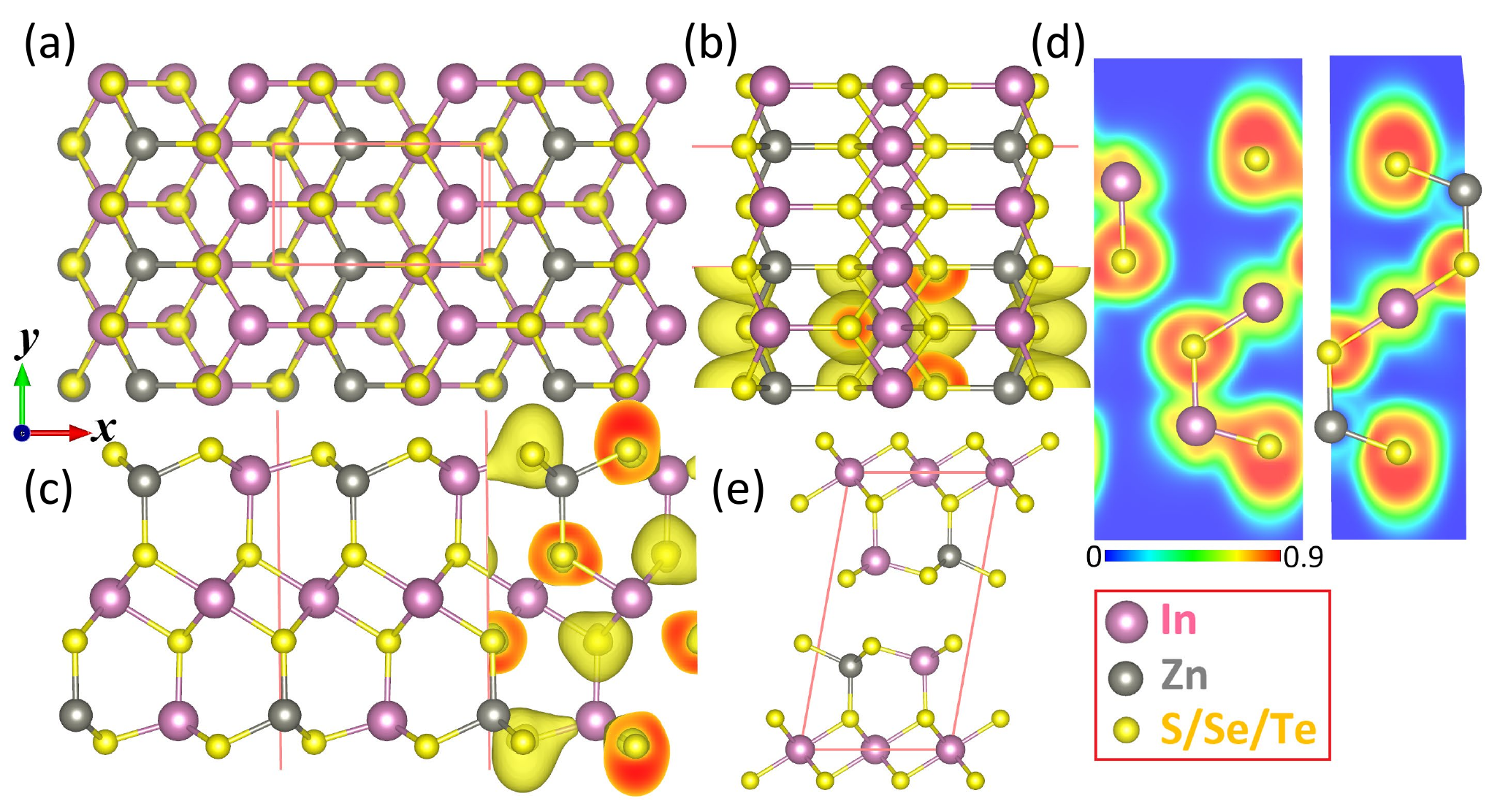}}
	\caption{Top and side views of the (a-d) single-layer and (e) bulk ZnIn$_2$S$_4$ along with the electron localization function (ELF) presented in the side views and sections. The ELF isosurface value is set to 0.75.}
	\label{Fig_struct}
\end{figure*}

First of all, we investigate the structural properties and the bonding mechanism of the ZnIn$_2$X$_4$ layered systems. We illustrate the optimized lattice structure of the single-layer and bulk ZnIn$_2$S$_4$ along with the isosurface and section maps of the electron localization function (ELF) in Fig.~\ref{Fig_struct}.
To highlight the anisotropicity in the transport response of the ZnIn$_2$X$_4$ monolayers, the two different directions of x and y are marked. The unit cell belongs to the triclinic crystal lattice and contains two Zn, four In, and eight chalcogen atoms, in which a 1T$-$InX$_2$ layer is sandwiched by the other two atomic layers. The lattice constants of the stress-free ZnIn$_2$S$_4$, ZnIn$_2$Se$_4$, and ZnIn$_2$Te$_4$ monolayers along the \mbox{x (y)} direction are \mbox{6.783 (3.925)}, 7.089 (4.100), and \mbox{7.625 (4.407) \AA}, respectively, which reveal that with increasing the atomic weight of the chalcogen atom, the lattice constants increase. Therefore, one can say that the lattice constants of the ZnIn$_2$Se$_4$ monolayer are almost the average value of those of the ZnIn$_2$S$_4$ and ZnIn$_2$Te$_4$ monolayers. The coordination of the optimized lattice structure of single-layer and bulk ZnIn$_2$X$_4$ are included in the supplementary information document.

The ELF is a spatial function between \mbox{0 and 1}. The ELF values close to unity reveal the strong covalent interaction or lone pair electrons, whereas lower values represent weaker ionic, metallic, or van der Waals interactions. From the ELF results, the existence of lone pairs is visible around the chalcogen atoms. Because of the higher electronegativity of chalcogen atoms than the Zn and In counterparts, they tend to attract electrons from their neighboring atoms. This explains the high electron localization around the chalcogen atoms, whereas Zn atoms are almost free of electron localization. From the ELF section and isosurface results, it is clear that the ELF values around the center of the X-In bonds are larger than 0.7, indicating the formation of covalent bonding. For the Zn-X bonds, the ELF around the center of bonds exhibits a sharp pattern with high values extending toward the chalcogen atom, and electron gas behavior toward the Zn atom, revealing the presence of ionic-type interactions along these bonds.

\begin{figure*}
	\centering{
	\includegraphics[width=0.97\textwidth]{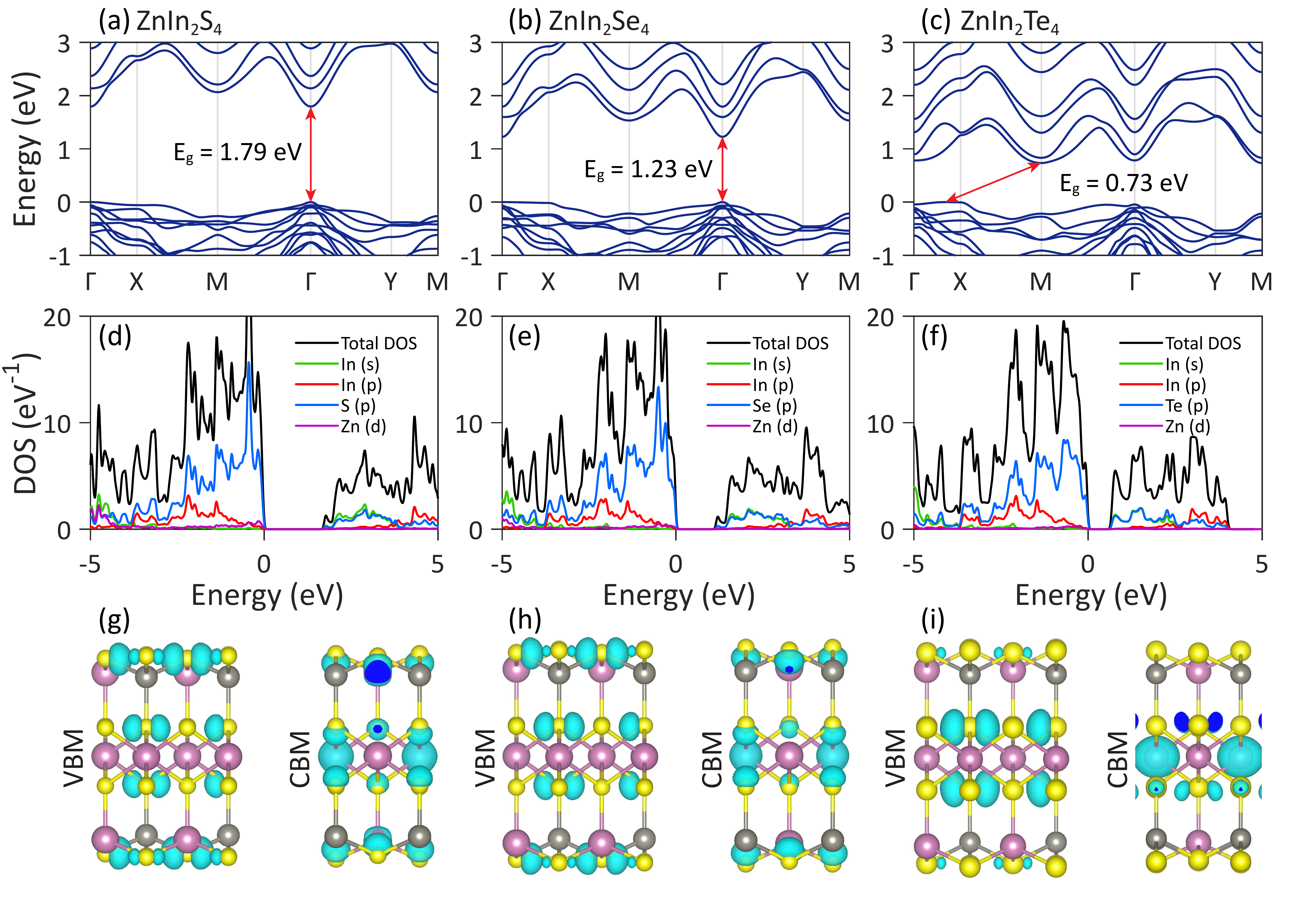}}
	\caption{Electronic properties of the ZnIn$_2$X$_4$ monolayers at the PBE level of theory. The total and partial density of states, and wave functions squared at the VBM and CBM for every monolayer are illustrated on the beneath panels, respectively. The VBMs are set to zero.}
	\label{Fig_band}
\end{figure*}

Fig.~\ref{Fig_band} represents the electronic properties of the ZnIn$_2$X$_4$ monolayers. As it is clear, the PBE functional predicts the monolayers to be semiconductors with band gaps of 1.79, 1.23, and
\mbox{0.73 eV} in such a way that the value of band gap decreases with the increase in the atomic weight of the chalcogen atom in the ZnIn$_2$X$_4$ monolayers. For the ZnIn$_2$S$_4$ and ZnIn$_2$Se$_4$, the valence band maximum (VBM) and the conduction band minimum (CBM) are positioned at the $\Gamma$ point, indicating a direct band gap while for the ZnIn$_2$Te$_4$, the VBM lies in the \mbox{$\Gamma-X$} direction, and the CBM is at the $M$ point, forming an indirect band gap. The conduction band edges are parabolically dispersed, showing free electrons. On the contrary, the valence band edges are flat, especially in the \mbox{$\Gamma-X$} direction, indicating strongly localized holes. Therefore, by regulating the Fermi level, one can have strongly localized holes and free electrons simultaneously in these monolayers, which could bring intriguing features like ferromagnetism and superconductivity \cite{mohebpour2020tuning}. Since the PBE functional mostly underestimates the electronic band gap \cite{moheb2022tr}, we employed the HSE06 hybrid functional with the default mixing parameter ($\alpha=0.25$). At this level, the band gaps are found to be 2.74, 1.98, and 1.28 eV for the ZnIn$_2$S$_4$, ZnIn$_2$Se$_4$, and ZnIn$_2$Te$_4$, respectively. The HSE06 band structures are similar to the PBE ones in such a way that the only difference is a rigid shift of the conduction band states with respect to the VBMs \mbox{(see Fig.~S1)}. Because the HSE06 functional does not include the electron-electron exchange interaction, we went beyond that by utilizing the single-shot G$_0$W$_0$ approach. At this level, the QP band gaps are predicted to be 3.94 (D), 2.77 (D), and \mbox{1.84 (I) eV} for the ZnIn$_2$S$_4$, ZnIn$_2$Se$_4$, and ZnIn$_2$Te$_4$, respectively, implying \mbox{self-energy} corrections of 2.15, 1.54, and \mbox{1.11 eV}. Indeed, with increasing the atomic weight of the chalcogen atom, the self-energy correction decreases. For comparison, the band gaps calculated with different approaches are tabulated in Table~\ref{gaps}.

\begin{table}
	\centering
	\normalsize
	\caption{Band gaps of the ZnIn$_2$X$_4$ monolayers calculated at different levels of theory in the unit of eV. The direct (D) and \mbox{indirect (I)} nature of the electronic band gaps is shown.}
	\label{gaps}
	\begin{tabular}{lcccccccc}
		\hline
		~Structure~~~&Type&~~PBE~~~&~~~HSE06~~~&~~~G$_0$W$_0$~~~&~~~Optical~~~\\
		\hline\hline
		~ZnIn$_2$S$_4$ &D&1.79 &	2.74  &	3.94   &   3.43 \\
		\hline
		~ZnIn$_2$Se$_4$&D&1.23 &	1.98  &	2.77   &   2.36 \\
		\hline
		~ZnIn$_2$Te$_4$&I&0.73 &	1.28  &	1.84   &   1.79 \\
		\hline
	\end{tabular}
\end{table}

As mentioned before, the bands are highly anisotropic along the $\Gamma-X$ and $\Gamma-Y$ directions due to the asymmetric crystal structures. To see this more clearly, we computed the effective mass of carriers. For the ZnIn$_2$S$_4$ monolayer, the effective masses of holes (electrons) are -4.47 (0.22) and \mbox{-0.29} (0.23) m$_0$ along the $\Gamma-X$ and $\Gamma-Y$ directions, respectively. Such a large effective mass of holes is the consequence of a flat valence band in the electronic structure. Also, for the ZnIn$_2$Se$_4$ monolayer, the corresponding effective masses are estimated to be -13.09 (0.17) and -0.24 (0.18) m$_0$, respectively. Therefore, one can conclude that with the increase in the atomic weight of chalcogen atom, the anisotropy increases. In the ZnIn$_2$Te$_4$ structure, the effective masses of holes reach -15.66 and -3.53 m$_0$ for the $VBM-X$ and $VBM-\Gamma$ directions, respectively. While, the effective masses of electrons become 0.93, 0.71, and 0.31 m$_0$ for the $M-X$, $M-\Gamma$, and $M-Y$ directions, respectively.

From Fig.~\ref{Fig_band} (d-f), it is also noticeable that the VBMs are mostly contributed by the \mbox{$p$-orbitals} of the chalcogen atoms while the CBMs are mainly dominated by the \mbox{$s$-orbitals} of the In atoms. It is also found that the density of states in valence bands is larger than that in the conduction bands. Therefore, we expect superior thermoelectric efficiency in the p-type doping as compared with the n-type counterpart. The participation of orbitals in the edges of bands can also be understood from the shapes of wave functions in real space, as illustrated in Fig.~\ref{Fig_band} (g-i). At the VBMs, the wave functions are shaped like dumbbells in such a way that they are distributed along the y-axis and centered on the chalcogen atoms, showing the \mbox{$p_y$-orbitals} of the chalcogen atoms. At the CBMs, the wave functions are spherically centered on the In atoms, representing the contribution of \mbox{$s$-orbitals} of these atoms.

\begin{figure*}
	\centering{
	\includegraphics[width=0.98\textwidth]{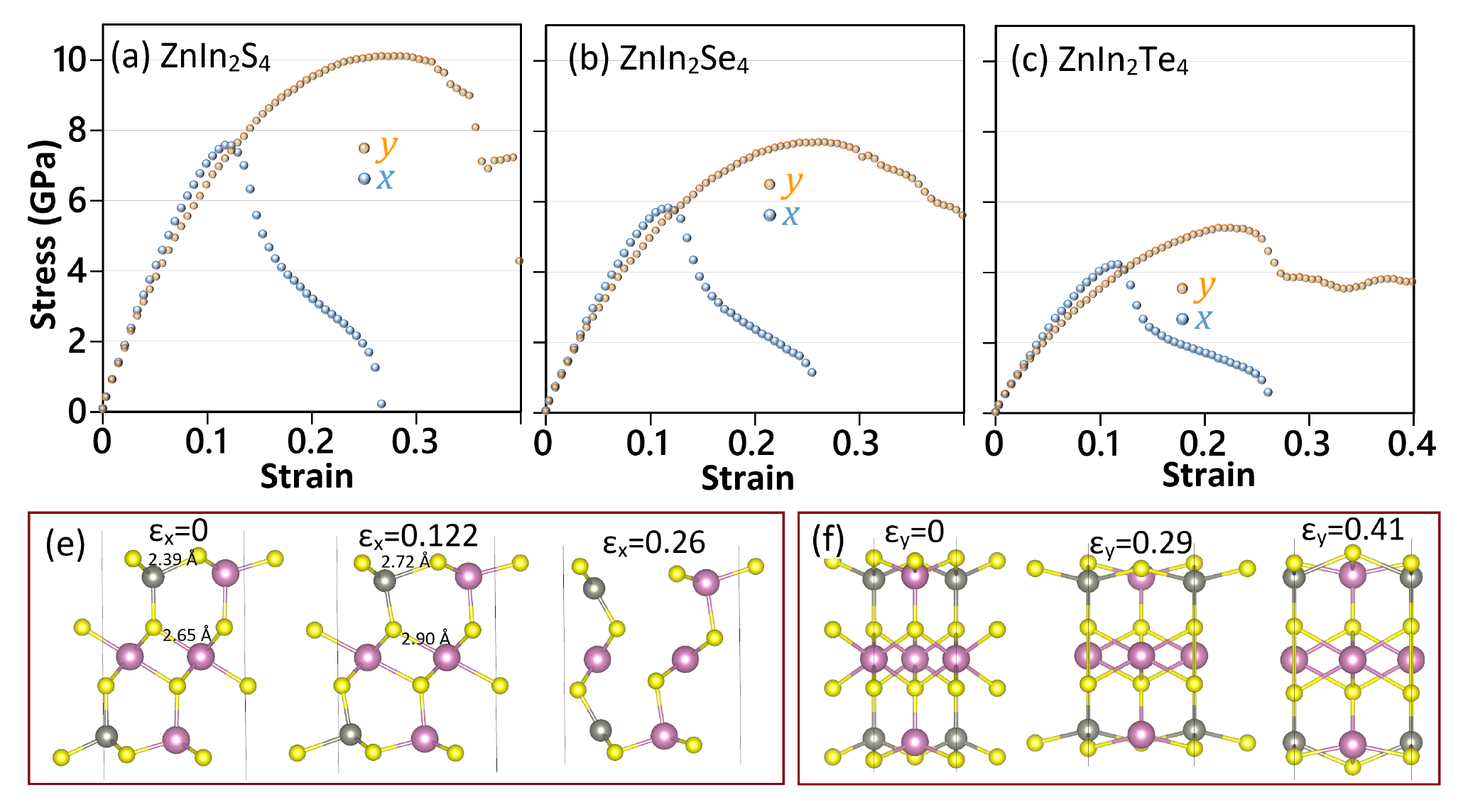}}
	\caption{(a-c) True uniaxial stress-strain relations of the ZnIn$_2$X$_4$ (X= S, Se, Te) monolayers elongated along the y and x directions. (e and f) present the stress-free and deformed ZnIn$_2$S$_4$ monolayer at different strain levels for the loading along the x ($\epsilon_x$) and y ($\epsilon_y$) directions.}
	\label{Fig_stress}
\end{figure*}

\subsection{Dynamical and Mechanical properties}
We now turn our attention to the analysis of the phononic and mechanical properties of the ZnIn$_2$X$_4$ monolayers. In this regard, we first study the phonon dispersion relations of the ZnIn$_2$X$_4$ monolayers obtained by the DFPT and MTP-based methods, as illustrated in Fig.~S4. Near the $\Gamma$ point, the out-of-plane acoustic modes (ZA) show quadratic relation for all the three considered monolayers, whereas the remaining two acoustic modes show linear dispersions \cite{chang2021origin, su2021investigation}. The  phonon dispersions confirm the dynamical stability of the ZnIn$_2$S$_4$, ZnIn$_2$Se$_4$, and ZnIn$_2$Te$_4$ monolayers because of the absence of imaginary phonon modes. Moreover, the comparison between the DFPT and MTP-based results confirm the remarkable accuracy of the developed classical models in reproducing the interatomic force constants.  It is clear that with the increase in the atomic weight of chalcogen atom, the dispersion of phonon modes in the entire frequency range shrinks considerably. The narrower dispersions of the phonon bands suggest the suppression of their corresponding group velocity, which most probably leads to a lower lattice thermal conductivity. This shrinkage also enhances the phonon band crossing, stimulating the higher scattering rates.
In addition, softening optical phonon modes with the increase in the atomic weight of chalcogen atom indicates loosening of bonds, which is in agreement with the ELF analysis.

We study the mechanical response of the ZnIn$_2$X$_4$ monolayers on the basis of uniaxial tensile results. In Fig.~\ref{Fig_stress}, the uniaxial stress-strain responses of the ZnIn$_2$S$_4$, ZnIn$_2$Se$_4$, and ZnIn$_2$Te$_4$ are compared. The predicted stress-strain relations are uniaxial and such that during the deformation, the structure is under stress only along the loading direction and is stress-free along the two other perpendicular directions. Since we study nanosheets that can freely move along their thickness direction, upon the geometry minimization, the system's stress component normal to the sheet naturally reaches a negligible value. Therefore, the cell size along the other in-plane perpendicular directions of the loading \mbox{(either x or y)} is adjusted to satisfy the negligible stress criteria after the geometry minimization. Note that the stress values are calculated at every strain by considering the real volume of the deformed monolayers. In this regard, the area of the monolayers can be easily obtained using the periodic simulation cell sizes along the planar direction. To calculate the area, the effective thickness at every step is calculated as the normal distance between boundary chalcogen atoms plus their effective van der Waals diameter (vdW). The thickness of the ZnIn$_2$S$_4$, ZnIn$_2$Se$_4$, and ZnIn$_2$Te$_4$ monolayers according to the geometry-optimized bulk lattices are predicted to be 12.434, 13.149, and 14.247 \AA, respectively. According to our geometry-optimized lattices, the normal distances between X-X atoms in the systems above are 9.633, 10.227, and 11.064 \AA, respectively, which are equivalent to the effective vdW diameters of 2.800, 2.922, and 3.182 \AA, for S, Se, and Te atoms, respectively, to satisfy the corresponding monolayers' thicknesses. The initial linear sections coincide closely for both considered loading directions, revealing a convincingly isotropic behavior. The elastic moduli of the ZnIn$_2$S$_4$, ZnIn$_2$Se$_4$, and ZnIn$_2$Te$_4$ along the \mbox{y- (x-) direction} are predicted to be 83 (84), 67 (63), and 47 (51) GPa, respectively. The ultimate tensile strength of the ZnIn$_2$S$_4$, ZnIn$_2$Se$_4$, and ZnIn$_2$Te$_4$ along the \mbox{y- (x-) direction} are predicted to be 10.1 (7.6), 7.7 (5.8), and 5.3 (4.2) GPa, respectively. These results reveal that the nanosheets are remarkably stronger along the y-direction than the x-direction, confirming their anisotropic tensile behavior despite an almost isotropic elasticity. The results confirm a clear reduction of the elastic modulus and tensile strength in the ZnIn$_2$X$_4$ nanosheets with the increase in the atomic weight of the chalcogen atom.

To better understand the anisotropic tensile strength in these nanosheets, in Fig.~\ref{Fig_stress} (e, f), we plot the deformed ZnIn$_2$S$_4$ monolayer at different strain levels. It appears that for the loading along the x-direction, the failure initiates along the Zn-S bonds, which are exactly oriented along the loading direction. It is worthwhile to remind that ELF results revealed the presence of ionic interactions along the aforementioned bonds. In contrast, for the loading along the y-direction, while no bonds are exactly aligned along the loading directions, but more bonds are inclined along the loading, and they consequently engage in the load transfer, resulting in higher tensile strengths.

\subsection{Optical properties}

\begin{figure*}
	\centering{
	\includegraphics[width=0.93\textwidth]{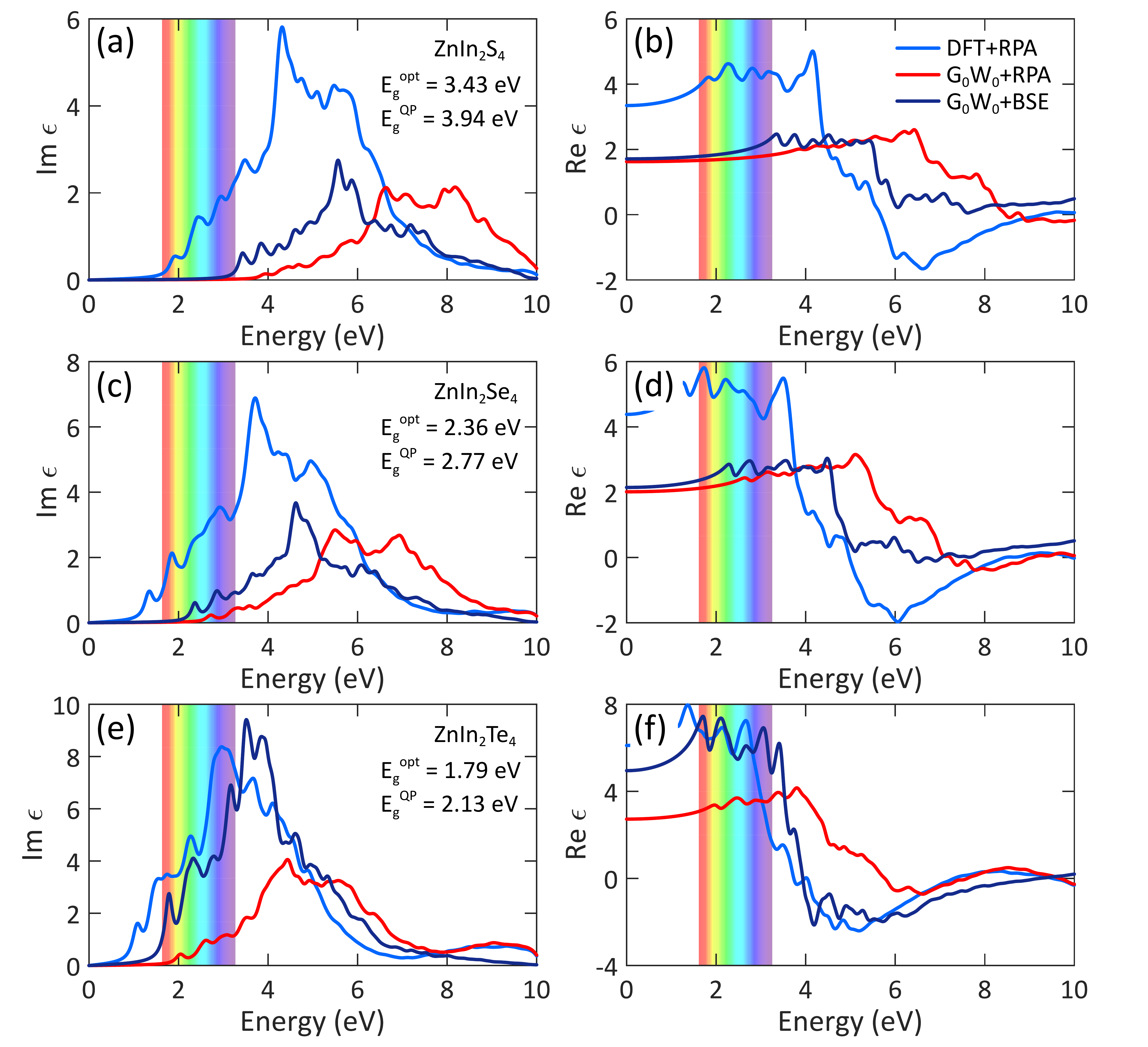}}
	\caption{Imaginary (left panel) and real (right panel) parts of the macroscopic dielectric functions of the ZnIn$_2$X$_4$ monolayers at three different levels of theory, namely DFT$+$RPA (without e$-$e and e$-$h interactions), G$_0$W$_0$$+$RPA (with e$-$e interaction, without e$-$h interaction), and G$_0$W$_0$$+$BSE (with e$-$e and e$-$h interactions). The optical gap and the G$_0$W$_0$ direct band gap of the monolayers are given. The visible light region is specified by a spectral color scheme.}
	\label{Fig_Im}
\end{figure*}

In this section, we study the optical properties of the ZnIn$_2$X$_4$ monolayers at distinguished levels of theory with and without the many-body effects, i.e. electron$-$electron and \mbox{electron$-$hole} interactions. We will see clearly that the aforementioned interactions dominate the optical response of the monolayers. Fig.~\ref{Fig_Im} (left panel) indicates the imaginary part of the dielectric functions of the ZnIn$_2$X$_4$ monolayers for the light polarized along the \mbox{x-direction}. Despite the different lattice constants along the x- and \mbox{y-direction}, the optical coefficients of the monolayers are almost isotropic. Therefore, we present only the results correlated with the polarization along the \mbox{x-direction}. Such an almost isotropic optical behavior is consistent with our analysis of elastic response. At the \mbox{DFT$+$RPA} level, as the simplest approximation, the imaginary parts of the dielectric functions of the monolayers are characterized by several peaks, which seem to have overestimated intensities owing to the exclusion of many-body effects. As can be seen, increasing the atomic weight of the chalcogen atom improves the intensity of the imaginary part. Including the electron$-$electron interaction
\mbox{(i.e. G$_0$W$_0$$+$RPA)} results in a blue shift in the optical spectra and a decrease in their intensities due to the self-energy correction. At this level, the first peak of optical spectrum appears at 3.94, 2.72, and \mbox{2.04 eV} for the ZnIn$_2$S$_4$, ZnIn$_2$Se$_4$, and ZnIn$_2$Te$_4$, respectively. These peaks correspond to direct transitions from the VBM, the \mbox{$p$-orbitals} of the chalcogen atoms, to the CBM, the \mbox{$s$-orbitals} of the In atoms. Taking into account the electron$-$hole interaction \mbox{(i.e. G$_0$W$_0$$+$BSE)} causes a cancellation effect and a red shift in the optical spectra of the monolayers in such a manner that their first peaks are located between the PBE and G$_0$W$_0$ band gaps at 3.43, 2.36, and \mbox{1.79 eV} for the ZnIn$_2$S$_4$, ZnIn$_2$Se$_4$, and ZnIn$_2$Te$_4$, respectively.
These peaks correspond to strongly bound bright excitons, which are referred to as Frenkel excitons. Accordingly, the exciton binding energies, the difference between the QP direct band gaps and the optical gaps, are 0.51, 0.41, and 0.34 eV for the ZnIn$_2$S$_4$, ZnIn$_2$Se$_4$, and ZnIn$_2$Te$_4$, respectively. It is clear that with the increase in the atomic weight of the chalcogen atom, the exciton binding energy decreases. However, such binding energies indicate the high stability of the excitonic states against thermal dissociation at \mbox{300 K}. They also show that the Coulomb interaction between the electron and hole forming the ground-state exciton is strong, hence, the exciton tends to be small of the same order as the size of the unit cell. Also, it is understood that increasing the atomic weight of the chalcogen atom leads to a redshift in the imaginary part of the dielectric function in such a manner that one can deduce that the ZnIn$_2$Te$_4$ monolayer is a very convenient candidate for optoelectronic applications in the visible area.

Aside from the similar impacts of many-body interactions on the real part of the dielectric function, from Fig.~\ref{Fig_Im} \mbox{(right panel)}, one can see that increasing the atomic weight of the chalcogen atom enhances the static dielectric constant of the monolayer. The obtained static value at the G$_0$W$_0$$+$BSE level is 1.71, 2.15, and 4.95 for the ZnIn$_2$S$_4$, ZnIn$_2$Se$_4$, and ZnIn$_2$Te$_4$ monolayers, respectively. Therefore, one can say that the larger the static dielectric constant, the smaller the exciton binding energy will be. Moreover, it is conspicuous that the spectra become negative in specific ranges, suggesting the metallic behavior of the monolayers at these ranges. Due to the considerable differences of the optical spectra with and without considering many-body effects, it can be concluded that these interactions play a vital role in the optical properties of the ZnIn$_2$X$_4$ monolayers.

The other optical coefficients of the monolayers obtained by the G$_0$W$_0$$+$BSE calculations are available in Fig.~S2. From Fig.~S2 (a), the static refractive indexes of the ZnIn$_2$S$_4$, ZnIn$_2$Se$_4$, and ZnIn$_2$Te$_4$ are found to be 1.31, 1.46, and 2.23 while the maximum refractive indexes are obtained 1.64, 1.89, and 2.85, respectively. After reaching the peaks, the refractive indexes drop progressively until they are less than that of the glass ($\sim$1.5), suggesting the high transparency of the monolayers at these ranges of energy. Importantly, for energies higher than \mbox{4 eV}, the one with the smallest band gap, the ZnIn$_2$Te$_4$ monolayer, manifests the smallest refractive index. Similarly, the extinction coefficients of the monolayers increase rapidly with increasing the photon energy; afterwards, they decrease gradually. From \mbox{Fig.~S2 (b)}, the maximum value of the extinction coefficient of the ZnIn$_2$S$_4$, ZnIn$_2$Se$_4$, and ZnIn$_2$Te$_4$ is found to be 0.93, 1.10, and 2.13 at 5.92, 4.92, and \mbox{4.11 eV}, respectively. This means that at these energies, the photons will be absorbed very fast.

As shown in Fig.~S2 (c), the absorption threshold \mbox{($\alpha>5\times10^{6}~$m$^{-1}$)} of the ZnIn$_2$S$_4$, ZnIn$_2$Se$_4$, and ZnIn$_2$Te$_4$ monolayers is located at 3.38, 2.37, and 1.71 eV, respectively, approximately where the imaginary part of the dielectric functions reaches the first peak. Importantly, the mean value of absorption coefficient in the visible area \mbox{(1.63 to 3.26 eV)} is 0.04, 0.37, and 1.86$\times$10$^7$~m$^{-1}$ for ZnIn$_2$S$_4$, ZnIn$_2$Se$_4$, and ZnIn$_2$Te$_4$, respectively. Therefore, it is conspicuous that the ZnIn$_2$Te$_4$ yields the highest absorption coefficient, which was expected considering our earlier finding of this monolayer, having the smallest optical gap. In addition, the maximum value of absorption for ZnIn$_2$S$_4$, ZnIn$_2$Se$_4$, and ZnIn$_2$Te$_4$ is found to be at 5.9, 6.4, and 5.7 eV, respectively. This means that the density of transitions at these particular energies is significant. As shown in \mbox{Fig.~S2 (d)}, one can manifest that increasing the atomic weight of the chalcogen atom increases reflectivity. More specifically, the mean value of reflectivity in the visible area is 2.8\%, 6.2\%, and 23.2\% for ZnIn$_2$S$_4$, ZnIn$_2$Se$_4$, and ZnIn$_2$Te$_4$, respectively. Overall, we predict the ZnIn$_2$Te$_4$ monolayer to be a potential candidate for optoelectronic applications.

Each peak in the absorption spectrum corresponds to at least one direct interband transition. To see the possibility of direct transitions from a specific valence band to a specific conduction band, we calculated the magnitude of transition dipole moment of the ZnIn$_2$X$_4$ monolayers. As represented in Fig.~S3 (a), the amplitude of transition dipole moment (TDM) is zero at the $M$, $\Gamma$, and $Y$ points for transitions from the highest valence band to the lowest conduction band \mbox{(i.e. V42 $\rightarrow$ C43)}. It is also zero for the \mbox{V41 $\rightarrow$ C43} transitions. This is why these transitions are named forbidden. However, the amplitude of TDM is non-zero ($\sim$230 Debye$^2$) at the $\Gamma$ point for the \mbox{V40 $\rightarrow$ C43} transition, which is why it is called allowed transition. \mbox{In general}, one can state that the most probable transitions occur near the $\Gamma$ point.

\subsection{Thermoelectric properties}

\begin{figure}
	\centering{
		\includegraphics[width=0.46\textwidth]{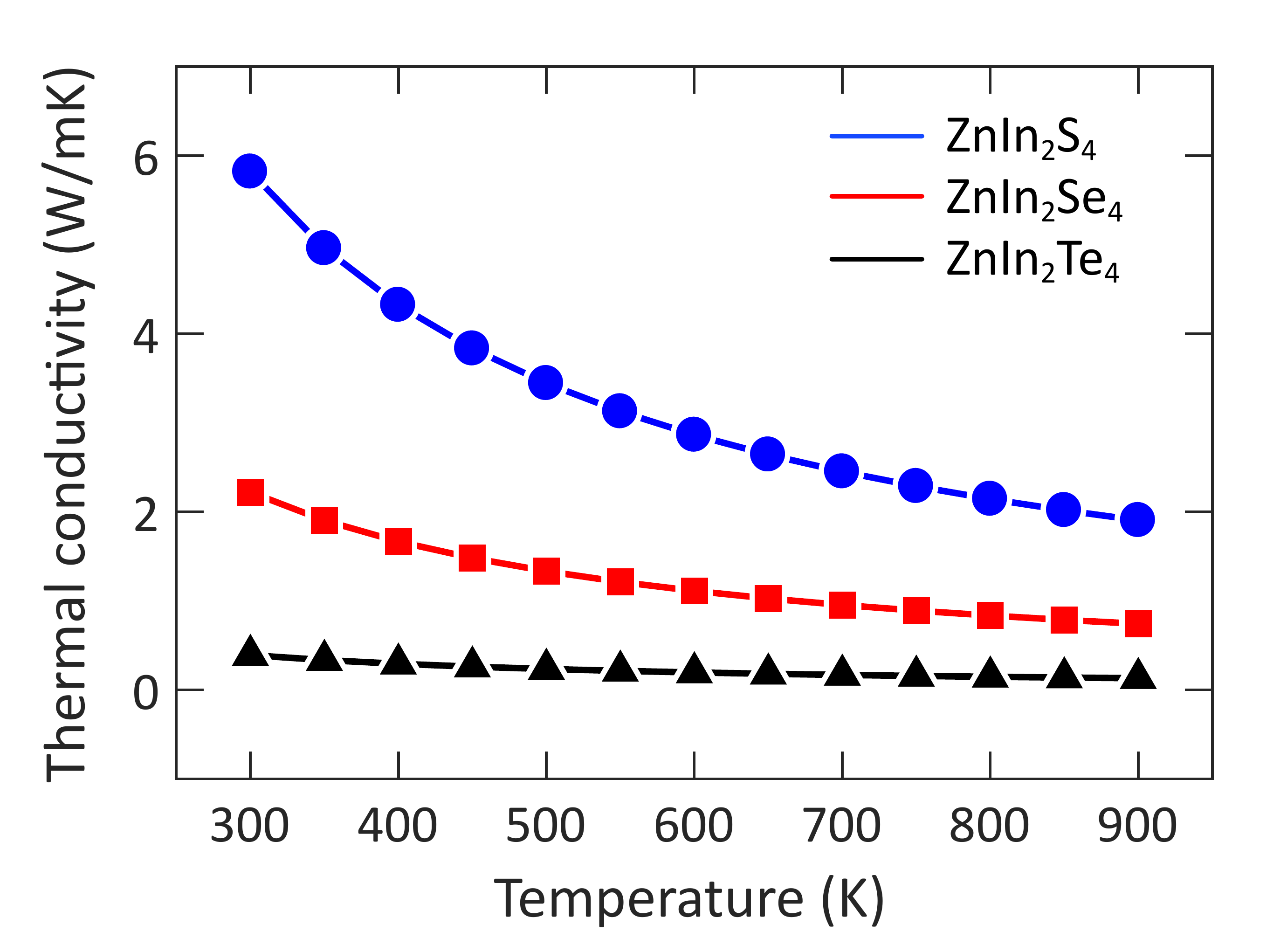}}
	\caption{Lattice thermal conductivity of the ZnIn$_2$X$_4$ monolayers as a function of temperature.}
	\label{Fig_conduct}
\end{figure}

We investigate the predicted temperature-dependent lattice thermal conductivity of the ZnIn$_2$S$_4$, ZnIn$_2$Se$_4$, and ZnIn$_2$Te$_4$ monolayers, as illustrated in Fig.~\ref{Fig_conduct}. In accordance with our analysis of the elastic response and optical properties, we found convincingly isotropic lattice thermal conductivity along these novel monolayers. The phononic thermal conductivity of the ZnIn$_2$S$_4$, ZnIn$_2$Se$_4$ and ZnIn$_2$Te$_4$, taking into account the isotope scattering at 300 K, are predicted to be remarkably low, 5.8, 2.0, and 0.4 W/mK, respectively. Normally, the lattice thermal conductivity follows a $\sim$$T^{-\lambda}$ trend with temperature (T), in which $\lambda$ is the temperature power factor. We estimate temperature power factors of 1.01, 1.0, and 1.0 for the ZnIn$_2$S$_4$, ZnIn$_2$Se$_4$, and ZnIn$_2$Te$_4$ monolayers, respectively. As expected, with the increase in the atomic weight of chalcogen atom, the lattice thermal conductivity decreases, which is also consistent with the classical theory saying a material with a lower elastic modulus and higher atomic weight yields a lower thermal conductivity.

To better understand the underlying mechanism resulting in the strong dependency of lattice thermal conductivity to the type of chalcogen atoms in these systems, in \mbox{Fig.~S5}, we compare the phonon's group velocity and lifetime of the ZnIn$_2$X$_4$ monolayers. As expected and shown in Fig.~S5 (a), with the increase in the atomic weight of chalcogen atom, the phonons' group velocities are clearly suppressed, that is consistent with the observed narrower dispersions for phonon modes and softer bonds. A similar observation is also found to be consistent for the phonons' lifetime illustrated in Fig.~S5 (b), which shows the substantial increase in the phonons scattering with the increase in the atomic weight of chalcogen atom. The maximum phonon group velocity in the ZnIn$_2$S$_4$, ZnIn$_2$Se$_4$, and ZnIn$_2$Te$_4$ monolayers is predicted to be 5.23, 5.54, and 3.68 km/s, respectively. As it is clear, the ZnIn$_2$Te$_4$ monolayer exhibits an ultralow thermal conductivity, which might be highly appealing for thermoelectric energy conversion.

\begin{figure*}
	\centering{
	\includegraphics[width=0.88\textwidth]{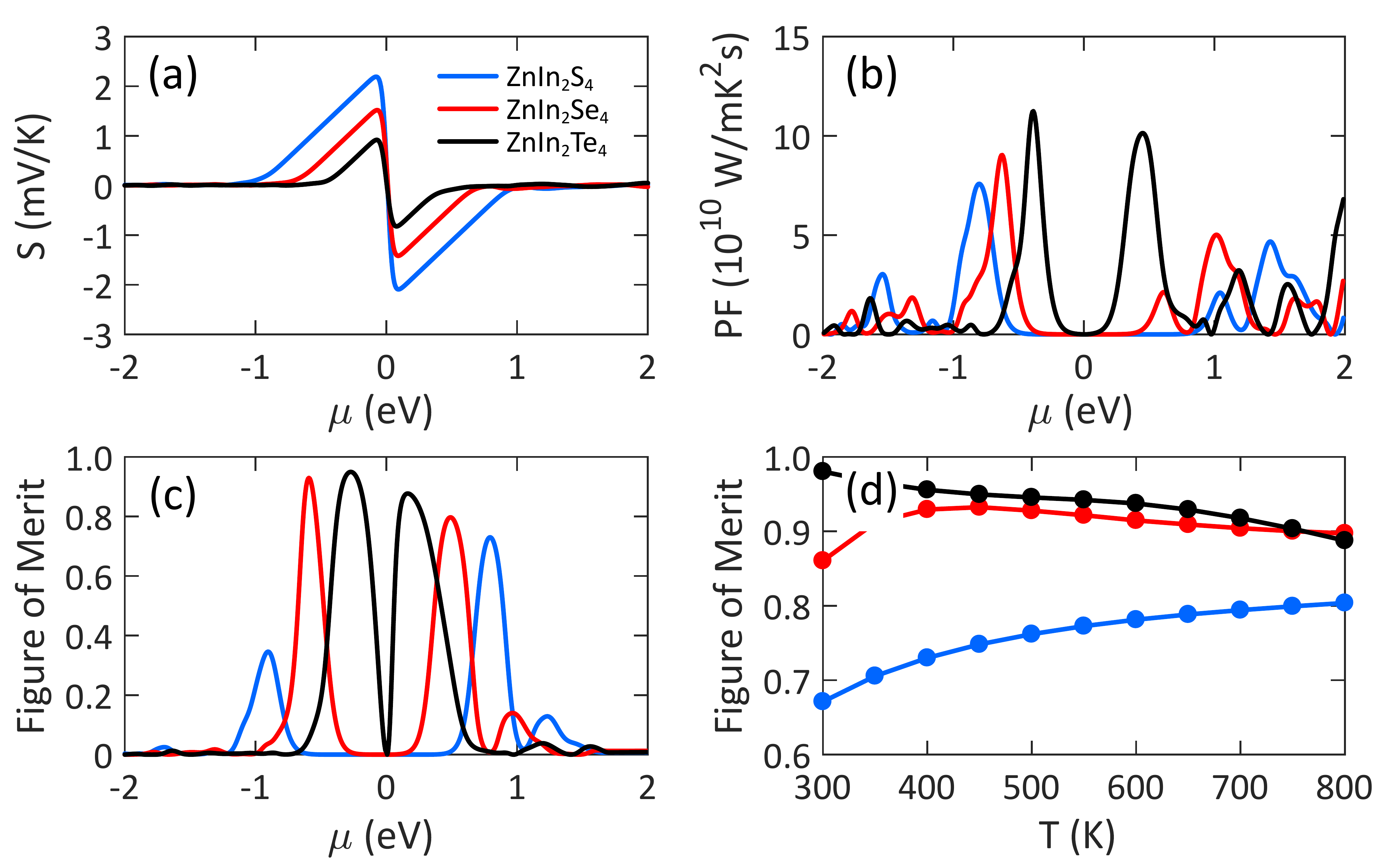}}
	\caption{(a) Thermopower, (b) power factor, and (c) figure of merit as a function of chemical potential for the considered monolayers at 400K. (d) Maximum of the figure of merit versus temperature for different structures.}
	\label{Fig_thermo}
\end{figure*}

The dependence of thermopower on chemical potential ($\mu$) for different structures is shown in Fig.~\ref{Fig_thermo} (a). The negative (positive) value of the chemical potential indicates p-type \mbox{(n-type)} doping. Large thermopower values are observed around the Fermi level on both sides, indicating that a doping of the order of \mbox{$10^{13}$~cm$^{-2}$} can achieve the optimal thermopower. Such a doping order can be easily obtained in the experiment using electric gates. The thermopower shows several dominant features, including a decrease of the thermopower with temperature as shown in Fig.~S6 for all the considered nanosheets, and a decrease of the thermopower with increasing the atomic weight of chalcogen atom because of the reduced band gap. Indeed, the magnitude of the thermopower is directly related to the band gap, and as we discussed earlier, the band gap in these systems reduces with increasing the atomic weight of the chalcogen atom. It is also noticeable that the thermopower vanished at energies far from the Fermi level due to the bi-particle effect. Moreover, the thermopower exhibits almost similar values for both \mbox{p-type} and \mbox{n-type} doping due to the symmetric nature of the valence and conduction bands. The maximum value of the thermopower at room temperature for different monolayers is tabulated in \mbox{Table \ref{table1}}.

To compute the relaxation time, we used the Bardeen–Shockley deformation potential \cite{bardeen1950deformation}, considering the coupling of acoustic phonons with electrons, as:
\begin{equation}
\label{time}
\tau=\frac{\hbar^3C^{2D}}{~k_BTm^*m_dE_l^2~},
\end{equation}
where $\hbar$, $C^{2D}$, $m_d$, $E_l$ denote the reduced Plank constant, elastic modulus, effective mass, and deformation potential, respectively. The obtained electron and hole relaxation times are listed in \mbox{Table \ref{table1}}. It is obvious that $\tau_e>\tau_h$, which is attributed to the smaller effective mass and deformation potential of the conduction band. Fig.~\ref{Fig_thermo} (b) shows the power factor, $PF=\frac{S^2\sigma}{\tau}$, as a function of the chemical potential for the considered monolayers. Unlike the thermopower, it is noticeable that the $PF$ increases with the temperature. In addition, its gap is directly dependent on the electronic band gap; hence, the gap is reduced with the increase in the atomic weight of the chalcogen atom. Furthermore, if we consider the anisotropy of the relaxation time for electrons and holes, the $PF$ of the n-doped monolayers is significantly higher than that of the p-doped system, as listed in \mbox{Table~\ref{table1}}.

\begin{table*}
	\normalsize
	\caption{Maximum of thermopower ($S_{max}$), hole and electron relaxation time, and maximum of $PF$ for the p- and n-type doping at 300 K.}
	\begin{tabular*}{\textwidth}{@{\extracolsep{\fill}}lllllllll}
		\hline
		Structure&$S_{max}^{h}$~(mV/K)&$S_{max}^{e}$ (mV/K)&$\tau_{h}$~($10^{-13}$s) & $\tau_{e}$~($10^{-13}$s)& $PF_{max}^h$~(mW/m.K)& $PF_{max}^e$~(mW/m.K) \\ \hline\hline
		ZnIn$_2$S$_4$&	2.88&	-2.60&	1.11&	85.5&	5.8&	620 \\ \hline
		ZnIn$_2$Se$_4$&	2.01&	-1.92&	2.24&	92.1&	18.9&	492 \\ \hline
		ZnIn$_2$Te$_4$&	1.22&	-1.12&	4.14&	37.1&	40&	218 \\ \hline
	\end{tabular*}
	\label{table1}
\end{table*}

The figure of merit, $ZT=\frac{S^2\sigma T}{\kappa_e+\kappa_l}$, is directly dependent on the $PF$ and inversely dependent on the thermal conductivity. The thermal conductivity is composed of the electronic ($\kappa_e$) and phononic ($\kappa_L$) parts. The electronic part of the thermal conductivity is computed using Wiedemann–Franz law, \mbox{$\kappa_e=L\sigma T$}, where $L=1.5\times10^{-8}$ W$\Omega$K$^{-2}$, is the Lorentz number. It is clear that the thermoelectric energy conversion efficiency of the considered monolayers is improved with the increase in the atomic weight of the chalcogen atom in these systems. As we discussed previously, the lattice thermal conductivity significantly reduces for larger chalcogen atoms, leading to an increase in the figure of merit. However, as observed in Fig.~\ref{Fig_thermo} (a), the thermopower is inversely dependent on the chalcogen atomic number, therefore, the difference in the final value of $ZT$ is not significant. Unlike ZnIn$_2$X$_4$ 
\mbox{(X= S, Se)}, the figure of merit of the ZnIn$_2$Te$_4$ decreases with increasing temperature, as shown in \mbox{Fig.~\ref{Fig_thermo} (d)}. The reducing trend is attributed to the reduction of the thermopower with temperature. The maximum $ZT$ of the predicted monolayers is close to unity, showing high thermoelectric performance. The thermopower and figure of merit of the ZnIn$_2$X$_4$ monolayers are higher than those for the similar structures. The maximum thermopower of $\alpha$-In$_2$Se$_3$ is 800~$\mu$V/K \cite{nian2021thermoelectric}, which is half of that predicted for ZnIn$_2$Se$_4$. The highest thermopower reported for Sn$_2$Bi is 300~$\mu$V/K \cite{mohebpour2020thermoelectric} while it is 500 $\mu$V/K for the Pd$_2$Se$_3$ monolayer \cite{naghavi2018pd2se3}. Also, the thermopower of WS$_2$ and WSTe monolayers are 328 and 322~$\mu$V/K \cite{patel2020high}, much lower than our results.

The figure of merit of the ZnIn$_2$X$_4$ monolayers as a function of chemical potential at different temperatures are provided in \mbox{Fig.~S7}. As can be found, for ZnIn$_2$S$_4$ monolayer, the figure of merit increases by 0.36 and reaches 0.69 at
\mbox{800 K}. While for the other ones, the figure of merit stays almost constant at 0.85. Overall, with ultralow lattice thermal conductivity, high thermopower, and large figure of merit, the ZnIn$_2$X$_4$ \mbox{(X= S, Se, Te)} monolayers can be promising candidates for the thermoelectric energy conversion systems.

\section{Conclusion}

We  conducted extensive first-principles calculations to explore the physical properties of the ZnIn$_2$X$_4$ \mbox{(X= S, Se, Te)} monolayers. We showed that the considered monolayers are dynamically and mechanically stable. We found that the ZnIn$_2$S$_4$, ZnIn$_2$Se$_4$, and ZnIn$_2$Te$_4$ monolayers are semiconductors with the G$_0$W$_0$ band gaps of \mbox{3.94 (D)}, \mbox{2.77 (D)}, and
\mbox{1.84 (I) eV}, respectively. We also observed strongly localized holes and free electrons in these 2D materials. The ZnIn$_2$Te$_4$ monolayer is found to yield a remarkably high absorption coefficient in the visible light region while its reflectivity rate remains less than 25\%, suggesting a great potential for optoelectronic applications. The elastic moduli of the ZnIn$_2$S$_4$, ZnIn$_2$Se$_4$, and ZnIn$_2$Te$_4$ monolayers along the y (x) direction are predicted to be\mbox{ 83 (84)}, 67 (63), and 47 (51)~GPa, respectively. The results reveal that the ZnIn$_2$X$_4$ nanosheets exhibit almost isotropic elastic, optical and lattice thermal transport responses while indicate highly anisotropic tensile behavior. The phononic thermal conductivity of the ZnIn$_2$S$_4$, ZnIn$_2$Se$_4$, and ZnIn$_2$Te$_4$ monolayers, taking into account the isotope scattering at 300 K, are predicted to be ultralow as 5.8, 2.0, and \mbox{0.4 W/mK}, respectively. The results also reveal a decrease in the elastic modulus, tensile strength, phonon group velocity, phonon lifetime, lattice thermal conductivity, and exciton binding energy with the increase in the atomic weight of chalcogen atom in the ZnIn$_2$X$_4$ nanosheets. The ZnIn$_2$Se$_4$ and ZnIn$_2$Te$_4$ monolayers are found to exhibit a figure of merit close to unity at 400 K, showing excellent thermoelectric performance. Our results provide an extensive vision concerning the critical physical properties of the 2D ZnIn$_2$X$_4$ \mbox{(X= S, Se, Te)} semiconductors and highlight their applications in optoelectronics and energy conversion systems.

\section*{Acknowledgment}

B.M and M. A. M contributed equally to this work. B.M. and X.Z. appreciate the funding by the Deutsche Forschungsgemeinschaft (DFG, German Research Foundation) under Germany’s Excellence Strategy within the Cluster of Excellence PhoenixD (EXC 2122, Project ID 390833453). B. M and T. R. are greatly thankful to the VEGAS cluster at Bauhaus University of Weimar for providing the computational resources. A.V.S. is supported by the Russian Science Foundation (Grant No 18-13-00479, https://rscf.ru/project/18-13-00479/).

\section*{Declaration of Interests}
The authors declare that they have no conflict of interest.
	
\bibliographystyle{apsrev4-2} % Tell bibtex which bibliography style to use
\bibliography{Ref}

\begin{thebibliography}{56}%
\makeatletter
\providecommand \@ifxundefined [1]{%
 \@ifx{#1\undefined}
}%
\providecommand \@ifnum [1]{%
 \ifnum #1\expandafter \@firstoftwo
 \else \expandafter \@secondoftwo
 \fi
}%
\providecommand \@ifx [1]{%
 \ifx #1\expandafter \@firstoftwo
 \else \expandafter \@secondoftwo
 \fi
}%
\providecommand \natexlab [1]{#1}%
\providecommand \enquote  [1]{``#1''}%
\providecommand \bibnamefont  [1]{#1}%
\providecommand \bibfnamefont [1]{#1}%
\providecommand \citenamefont [1]{#1}%
\providecommand \href@noop [0]{\@secondoftwo}%
\providecommand \href [0]{\begingroup \@sanitize@url \@href}%
\providecommand \@href[1]{\@@startlink{#1}\@@href}%
\providecommand \@@href[1]{\endgroup#1\@@endlink}%
\providecommand \@sanitize@url [0]{\catcode `\\12\catcode `\$12\catcode
  `\&12\catcode `\#12\catcode `\^12\catcode `\_12\catcode `\%12\relax}%
\providecommand \@@startlink[1]{}%
\providecommand \@@endlink[0]{}%
\providecommand \url  [0]{\begingroup\@sanitize@url \@url }%
\providecommand \@url [1]{\endgroup\@href {#1}{\urlprefix }}%
\providecommand \urlprefix  [0]{URL }%
\providecommand \Eprint [0]{\href }%
\providecommand \doibase [0]{https://doi.org/}%
\providecommand \selectlanguage [0]{\@gobble}%
\providecommand \bibinfo  [0]{\@secondoftwo}%
\providecommand \bibfield  [0]{\@secondoftwo}%
\providecommand \translation [1]{[#1]}%
\providecommand \BibitemOpen [0]{}%
\providecommand \bibitemStop [0]{}%
\providecommand \bibitemNoStop [0]{.\EOS\space}%
\providecommand \EOS [0]{\spacefactor3000\relax}%
\providecommand \BibitemShut  [1]{\csname bibitem#1\endcsname}%
\let\auto@bib@innerbib\@empty
%</preamble>
\bibitem [{\citenamefont {Novoselov}\ \emph {et~al.}(2004)\citenamefont
  {Novoselov}, \citenamefont {Geim}, \citenamefont {Morozov}, \citenamefont
  {Jiang}, \citenamefont {Zhang}, \citenamefont {Dubonos}, \citenamefont
  {Grigorieva},\ and\ \citenamefont {Firsov}}]{novoselov2004electric}%
  \BibitemOpen
  \bibfield  {author} {\bibinfo {author} {\bibfnamefont {K.~S.}\ \bibnamefont
  {Novoselov}}, \bibinfo {author} {\bibfnamefont {A.~K.}\ \bibnamefont {Geim}},
  \bibinfo {author} {\bibfnamefont {S.~V.}\ \bibnamefont {Morozov}}, \bibinfo
  {author} {\bibfnamefont {D.-e.}\ \bibnamefont {Jiang}}, \bibinfo {author}
  {\bibfnamefont {Y.}~\bibnamefont {Zhang}}, \bibinfo {author} {\bibfnamefont
  {S.~V.}\ \bibnamefont {Dubonos}}, \bibinfo {author} {\bibfnamefont {I.~V.}\
  \bibnamefont {Grigorieva}},\ and\ \bibinfo {author} {\bibfnamefont {A.~A.}\
  \bibnamefont {Firsov}},\ }\href@noop {} {\bibfield  {journal} {\bibinfo
  {journal} {science}\ }\textbf {\bibinfo {volume} {306}},\ \bibinfo {pages}
  {666} (\bibinfo {year} {2004})}\BibitemShut {NoStop}%
\bibitem [{\citenamefont {Novoselov}\ and\ \citenamefont
  {Geim}(2007)}]{novoselov2007rise}%
  \BibitemOpen
  \bibfield  {author} {\bibinfo {author} {\bibfnamefont {K.~S.}\ \bibnamefont
  {Novoselov}}\ and\ \bibinfo {author} {\bibfnamefont {A.}~\bibnamefont
  {Geim}},\ }\href@noop {} {\bibfield  {journal} {\bibinfo  {journal} {Nat.
  Mater}\ }\textbf {\bibinfo {volume} {6}},\ \bibinfo {pages} {183} (\bibinfo
  {year} {2007})}\BibitemShut {NoStop}%
\bibitem [{\citenamefont {Neto}\ \emph {et~al.}(2009)\citenamefont {Neto},
  \citenamefont {Guinea}, \citenamefont {Peres}, \citenamefont {Novoselov},\
  and\ \citenamefont {Geim}}]{neto2009electronic}%
  \BibitemOpen
  \bibfield  {author} {\bibinfo {author} {\bibfnamefont {A.~C.}\ \bibnamefont
  {Neto}}, \bibinfo {author} {\bibfnamefont {F.}~\bibnamefont {Guinea}},
  \bibinfo {author} {\bibfnamefont {N.~M.}\ \bibnamefont {Peres}}, \bibinfo
  {author} {\bibfnamefont {K.~S.}\ \bibnamefont {Novoselov}},\ and\ \bibinfo
  {author} {\bibfnamefont {A.~K.}\ \bibnamefont {Geim}},\ }\href@noop {}
  {\bibfield  {journal} {\bibinfo  {journal} {Reviews of modern physics}\
  }\textbf {\bibinfo {volume} {81}},\ \bibinfo {pages} {109} (\bibinfo {year}
  {2009})}\BibitemShut {NoStop}%
\bibitem [{\citenamefont {Lee}\ \emph {et~al.}(2008)\citenamefont {Lee},
  \citenamefont {Wei}, \citenamefont {Kysar},\ and\ \citenamefont
  {Hone}}]{lee2008measurement}%
  \BibitemOpen
  \bibfield  {author} {\bibinfo {author} {\bibfnamefont {C.}~\bibnamefont
  {Lee}}, \bibinfo {author} {\bibfnamefont {X.}~\bibnamefont {Wei}}, \bibinfo
  {author} {\bibfnamefont {J.~W.}\ \bibnamefont {Kysar}},\ and\ \bibinfo
  {author} {\bibfnamefont {J.}~\bibnamefont {Hone}},\ }\href@noop {} {\bibfield
   {journal} {\bibinfo  {journal} {science}\ }\textbf {\bibinfo {volume}
  {321}},\ \bibinfo {pages} {385} (\bibinfo {year} {2008})}\BibitemShut
  {NoStop}%
\bibitem [{\citenamefont {Banszerus}\ \emph {et~al.}(2015)\citenamefont
  {Banszerus}, \citenamefont {Schmitz}, \citenamefont {Engels}, \citenamefont
  {Dauber}, \citenamefont {Oellers}, \citenamefont {Haupt}, \citenamefont
  {Watanabe}, \citenamefont {Taniguchi}, \citenamefont {Beschoten},\ and\
  \citenamefont {Stampfer}}]{banszerus2015ultrahigh}%
  \BibitemOpen
  \bibfield  {author} {\bibinfo {author} {\bibfnamefont {L.}~\bibnamefont
  {Banszerus}}, \bibinfo {author} {\bibfnamefont {M.}~\bibnamefont {Schmitz}},
  \bibinfo {author} {\bibfnamefont {S.}~\bibnamefont {Engels}}, \bibinfo
  {author} {\bibfnamefont {J.}~\bibnamefont {Dauber}}, \bibinfo {author}
  {\bibfnamefont {M.}~\bibnamefont {Oellers}}, \bibinfo {author} {\bibfnamefont
  {F.}~\bibnamefont {Haupt}}, \bibinfo {author} {\bibfnamefont
  {K.}~\bibnamefont {Watanabe}}, \bibinfo {author} {\bibfnamefont
  {T.}~\bibnamefont {Taniguchi}}, \bibinfo {author} {\bibfnamefont
  {B.}~\bibnamefont {Beschoten}},\ and\ \bibinfo {author} {\bibfnamefont
  {C.}~\bibnamefont {Stampfer}},\ }\href@noop {} {\bibfield  {journal}
  {\bibinfo  {journal} {Science advances}\ }\textbf {\bibinfo {volume} {1}},\
  \bibinfo {pages} {e1500222} (\bibinfo {year} {2015})}\BibitemShut {NoStop}%
\bibitem [{\citenamefont {Ghosh}\ \emph {et~al.}(2008)\citenamefont {Ghosh},
  \citenamefont {Calizo}, \citenamefont {Teweldebrhan}, \citenamefont
  {Pokatilov}, \citenamefont {Nika}, \citenamefont {Balandin}, \citenamefont
  {Bao}, \citenamefont {Miao},\ and\ \citenamefont {Lau}}]{ghosh2008extremely}%
  \BibitemOpen
  \bibfield  {author} {\bibinfo {author} {\bibfnamefont {D.}~\bibnamefont
  {Ghosh}}, \bibinfo {author} {\bibfnamefont {I.}~\bibnamefont {Calizo}},
  \bibinfo {author} {\bibfnamefont {D.}~\bibnamefont {Teweldebrhan}}, \bibinfo
  {author} {\bibfnamefont {E.~P.}\ \bibnamefont {Pokatilov}}, \bibinfo {author}
  {\bibfnamefont {D.~L.}\ \bibnamefont {Nika}}, \bibinfo {author}
  {\bibfnamefont {A.~A.}\ \bibnamefont {Balandin}}, \bibinfo {author}
  {\bibfnamefont {W.}~\bibnamefont {Bao}}, \bibinfo {author} {\bibfnamefont
  {F.}~\bibnamefont {Miao}},\ and\ \bibinfo {author} {\bibfnamefont {C.~N.}\
  \bibnamefont {Lau}},\ }\href@noop {} {\bibfield  {journal} {\bibinfo
  {journal} {Applied Physics Letters}\ }\textbf {\bibinfo {volume} {92}},\
  \bibinfo {pages} {151911} (\bibinfo {year} {2008})}\BibitemShut {NoStop}%
\bibitem [{\citenamefont {Balandin}\ \emph {et~al.}(2008)\citenamefont
  {Balandin}, \citenamefont {Ghosh}, \citenamefont {Bao}, \citenamefont
  {Calizo}, \citenamefont {Teweldebrhan}, \citenamefont {Miao},\ and\
  \citenamefont {Lau}}]{balandin2008superior}%
  \BibitemOpen
  \bibfield  {author} {\bibinfo {author} {\bibfnamefont {A.~A.}\ \bibnamefont
  {Balandin}}, \bibinfo {author} {\bibfnamefont {S.}~\bibnamefont {Ghosh}},
  \bibinfo {author} {\bibfnamefont {W.}~\bibnamefont {Bao}}, \bibinfo {author}
  {\bibfnamefont {I.}~\bibnamefont {Calizo}}, \bibinfo {author} {\bibfnamefont
  {D.}~\bibnamefont {Teweldebrhan}}, \bibinfo {author} {\bibfnamefont
  {F.}~\bibnamefont {Miao}},\ and\ \bibinfo {author} {\bibfnamefont {C.~N.}\
  \bibnamefont {Lau}},\ }\href@noop {} {\bibfield  {journal} {\bibinfo
  {journal} {Nano letters}\ }\textbf {\bibinfo {volume} {8}},\ \bibinfo {pages}
  {902} (\bibinfo {year} {2008})}\BibitemShut {NoStop}%
\bibitem [{\citenamefont {Berger}\ \emph {et~al.}(2004)\citenamefont {Berger},
  \citenamefont {Song}, \citenamefont {Li}, \citenamefont {Li}, \citenamefont
  {Ogbazghi}, \citenamefont {Feng}, \citenamefont {Dai}, \citenamefont
  {Marchenkov}, \citenamefont {Conrad}, \citenamefont {First} \emph
  {et~al.}}]{berger2004ultrathin}%
  \BibitemOpen
  \bibfield  {author} {\bibinfo {author} {\bibfnamefont {C.}~\bibnamefont
  {Berger}}, \bibinfo {author} {\bibfnamefont {Z.}~\bibnamefont {Song}},
  \bibinfo {author} {\bibfnamefont {T.}~\bibnamefont {Li}}, \bibinfo {author}
  {\bibfnamefont {X.}~\bibnamefont {Li}}, \bibinfo {author} {\bibfnamefont
  {A.~Y.}\ \bibnamefont {Ogbazghi}}, \bibinfo {author} {\bibfnamefont
  {R.}~\bibnamefont {Feng}}, \bibinfo {author} {\bibfnamefont {Z.}~\bibnamefont
  {Dai}}, \bibinfo {author} {\bibfnamefont {A.~N.}\ \bibnamefont {Marchenkov}},
  \bibinfo {author} {\bibfnamefont {E.~H.}\ \bibnamefont {Conrad}}, \bibinfo
  {author} {\bibfnamefont {P.~N.}\ \bibnamefont {First}}, \emph {et~al.},\
  }\href@noop {} {\bibfield  {journal} {\bibinfo  {journal} {The Journal of
  Physical Chemistry B}\ }\textbf {\bibinfo {volume} {108}},\ \bibinfo {pages}
  {19912} (\bibinfo {year} {2004})}\BibitemShut {NoStop}%
\bibitem [{\citenamefont {Liu}\ \emph {et~al.}(2011)\citenamefont {Liu},
  \citenamefont {Yin}, \citenamefont {Ulin-Avila}, \citenamefont {Geng},
  \citenamefont {Zentgraf}, \citenamefont {Ju}, \citenamefont {Wang},\ and\
  \citenamefont {Zhang}}]{liu2011graphene}%
  \BibitemOpen
  \bibfield  {author} {\bibinfo {author} {\bibfnamefont {M.}~\bibnamefont
  {Liu}}, \bibinfo {author} {\bibfnamefont {X.}~\bibnamefont {Yin}}, \bibinfo
  {author} {\bibfnamefont {E.}~\bibnamefont {Ulin-Avila}}, \bibinfo {author}
  {\bibfnamefont {B.}~\bibnamefont {Geng}}, \bibinfo {author} {\bibfnamefont
  {T.}~\bibnamefont {Zentgraf}}, \bibinfo {author} {\bibfnamefont
  {L.}~\bibnamefont {Ju}}, \bibinfo {author} {\bibfnamefont {F.}~\bibnamefont
  {Wang}},\ and\ \bibinfo {author} {\bibfnamefont {X.}~\bibnamefont {Zhang}},\
  }\href@noop {} {\bibfield  {journal} {\bibinfo  {journal} {Nature}\ }\textbf
  {\bibinfo {volume} {474}},\ \bibinfo {pages} {64} (\bibinfo {year}
  {2011})}\BibitemShut {NoStop}%
\bibitem [{\citenamefont {Withers}\ \emph {et~al.}(2010)\citenamefont
  {Withers}, \citenamefont {Dubois},\ and\ \citenamefont
  {Savchenko}}]{withers2010electron}%
  \BibitemOpen
  \bibfield  {author} {\bibinfo {author} {\bibfnamefont {F.}~\bibnamefont
  {Withers}}, \bibinfo {author} {\bibfnamefont {M.}~\bibnamefont {Dubois}},\
  and\ \bibinfo {author} {\bibfnamefont {A.~K.}\ \bibnamefont {Savchenko}},\
  }\href@noop {} {\bibfield  {journal} {\bibinfo  {journal} {Physical review
  B}\ }\textbf {\bibinfo {volume} {82}},\ \bibinfo {pages} {073403} (\bibinfo
  {year} {2010})}\BibitemShut {NoStop}%
\bibitem [{\citenamefont {Liu}\ and\ \citenamefont
  {Zhou}(2019)}]{liu2019recent}%
  \BibitemOpen
  \bibfield  {author} {\bibinfo {author} {\bibfnamefont {B.}~\bibnamefont
  {Liu}}\ and\ \bibinfo {author} {\bibfnamefont {K.}~\bibnamefont {Zhou}},\
  }\href@noop {} {\bibfield  {journal} {\bibinfo  {journal} {Progress in
  Materials Science}\ }\textbf {\bibinfo {volume} {100}},\ \bibinfo {pages}
  {99} (\bibinfo {year} {2019})}\BibitemShut {NoStop}%
\bibitem [{\citenamefont {Baughman}\ \emph {et~al.}(1987)\citenamefont
  {Baughman}, \citenamefont {Eckhardt},\ and\ \citenamefont
  {Kertesz}}]{baughman1987structure}%
  \BibitemOpen
  \bibfield  {author} {\bibinfo {author} {\bibfnamefont {R.}~\bibnamefont
  {Baughman}}, \bibinfo {author} {\bibfnamefont {H.}~\bibnamefont {Eckhardt}},\
  and\ \bibinfo {author} {\bibfnamefont {M.}~\bibnamefont {Kertesz}},\
  }\href@noop {} {\bibfield  {journal} {\bibinfo  {journal} {The Journal of
  chemical physics}\ }\textbf {\bibinfo {volume} {87}},\ \bibinfo {pages}
  {6687} (\bibinfo {year} {1987})}\BibitemShut {NoStop}%
\bibitem [{\citenamefont {Bakharev}\ \emph {et~al.}(2020)\citenamefont
  {Bakharev}, \citenamefont {Huang}, \citenamefont {Saxena}, \citenamefont
  {Lee}, \citenamefont {Joo}, \citenamefont {Park}, \citenamefont {Dong},
  \citenamefont {Camacho-Mojica}, \citenamefont {Jin}, \citenamefont {Kwon}
  \emph {et~al.}}]{bakharev2020chemically}%
  \BibitemOpen
  \bibfield  {author} {\bibinfo {author} {\bibfnamefont {P.~V.}\ \bibnamefont
  {Bakharev}}, \bibinfo {author} {\bibfnamefont {M.}~\bibnamefont {Huang}},
  \bibinfo {author} {\bibfnamefont {M.}~\bibnamefont {Saxena}}, \bibinfo
  {author} {\bibfnamefont {S.~W.}\ \bibnamefont {Lee}}, \bibinfo {author}
  {\bibfnamefont {S.~H.}\ \bibnamefont {Joo}}, \bibinfo {author} {\bibfnamefont
  {S.~O.}\ \bibnamefont {Park}}, \bibinfo {author} {\bibfnamefont
  {J.}~\bibnamefont {Dong}}, \bibinfo {author} {\bibfnamefont {D.~C.}\
  \bibnamefont {Camacho-Mojica}}, \bibinfo {author} {\bibfnamefont
  {S.}~\bibnamefont {Jin}}, \bibinfo {author} {\bibfnamefont {Y.}~\bibnamefont
  {Kwon}}, \emph {et~al.},\ }\href@noop {} {\bibfield  {journal} {\bibinfo
  {journal} {Nature nanotechnology}\ }\textbf {\bibinfo {volume} {15}},\
  \bibinfo {pages} {59} (\bibinfo {year} {2020})}\BibitemShut {NoStop}%
\bibitem [{\citenamefont {Wehling}\ \emph {et~al.}(2008)\citenamefont
  {Wehling}, \citenamefont {Novoselov}, \citenamefont {Morozov}, \citenamefont
  {Vdovin}, \citenamefont {Katsnelson}, \citenamefont {Geim},\ and\
  \citenamefont {Lichtenstein}}]{wehling2008molecular}%
  \BibitemOpen
  \bibfield  {author} {\bibinfo {author} {\bibfnamefont {T.}~\bibnamefont
  {Wehling}}, \bibinfo {author} {\bibfnamefont {K.}~\bibnamefont {Novoselov}},
  \bibinfo {author} {\bibfnamefont {S.}~\bibnamefont {Morozov}}, \bibinfo
  {author} {\bibfnamefont {E.}~\bibnamefont {Vdovin}}, \bibinfo {author}
  {\bibfnamefont {M.}~\bibnamefont {Katsnelson}}, \bibinfo {author}
  {\bibfnamefont {A.}~\bibnamefont {Geim}},\ and\ \bibinfo {author}
  {\bibfnamefont {A.}~\bibnamefont {Lichtenstein}},\ }\href@noop {} {\bibfield
  {journal} {\bibinfo  {journal} {Nano letters}\ }\textbf {\bibinfo {volume}
  {8}},\ \bibinfo {pages} {173} (\bibinfo {year} {2008})}\BibitemShut {NoStop}%
\bibitem [{\citenamefont {Schedin}\ \emph {et~al.}(2007)\citenamefont
  {Schedin}, \citenamefont {Geim}, \citenamefont {Morozov}, \citenamefont
  {Hill}, \citenamefont {Blake}, \citenamefont {Katsnelson},\ and\
  \citenamefont {Novoselov}}]{schedin2007detection}%
  \BibitemOpen
  \bibfield  {author} {\bibinfo {author} {\bibfnamefont {F.}~\bibnamefont
  {Schedin}}, \bibinfo {author} {\bibfnamefont {A.~K.}\ \bibnamefont {Geim}},
  \bibinfo {author} {\bibfnamefont {S.~V.}\ \bibnamefont {Morozov}}, \bibinfo
  {author} {\bibfnamefont {E.}~\bibnamefont {Hill}}, \bibinfo {author}
  {\bibfnamefont {P.}~\bibnamefont {Blake}}, \bibinfo {author} {\bibfnamefont
  {M.}~\bibnamefont {Katsnelson}},\ and\ \bibinfo {author} {\bibfnamefont
  {K.~S.}\ \bibnamefont {Novoselov}},\ }\href@noop {} {\bibfield  {journal}
  {\bibinfo  {journal} {Nature materials}\ }\textbf {\bibinfo {volume} {6}},\
  \bibinfo {pages} {652} (\bibinfo {year} {2007})}\BibitemShut {NoStop}%
\bibitem [{\citenamefont {Soriano}\ \emph {et~al.}(2015)\citenamefont
  {Soriano}, \citenamefont {Van~Tuan}, \citenamefont {Dubois}, \citenamefont
  {Gmitra}, \citenamefont {Cummings}, \citenamefont {Kochan}, \citenamefont
  {Ortmann}, \citenamefont {Charlier}, \citenamefont {Fabian},\ and\
  \citenamefont {Roche}}]{soriano2015spin}%
  \BibitemOpen
  \bibfield  {author} {\bibinfo {author} {\bibfnamefont {D.}~\bibnamefont
  {Soriano}}, \bibinfo {author} {\bibfnamefont {D.}~\bibnamefont {Van~Tuan}},
  \bibinfo {author} {\bibfnamefont {S.~M.}\ \bibnamefont {Dubois}}, \bibinfo
  {author} {\bibfnamefont {M.}~\bibnamefont {Gmitra}}, \bibinfo {author}
  {\bibfnamefont {A.~W.}\ \bibnamefont {Cummings}}, \bibinfo {author}
  {\bibfnamefont {D.}~\bibnamefont {Kochan}}, \bibinfo {author} {\bibfnamefont
  {F.}~\bibnamefont {Ortmann}}, \bibinfo {author} {\bibfnamefont {J.-C.}\
  \bibnamefont {Charlier}}, \bibinfo {author} {\bibfnamefont {J.}~\bibnamefont
  {Fabian}},\ and\ \bibinfo {author} {\bibfnamefont {S.}~\bibnamefont
  {Roche}},\ }\href@noop {} {\bibfield  {journal} {\bibinfo  {journal} {2D
  Materials}\ }\textbf {\bibinfo {volume} {2}},\ \bibinfo {pages} {022002}
  (\bibinfo {year} {2015})}\BibitemShut {NoStop}%
\bibitem [{\citenamefont {Lherbier}\ \emph {et~al.}(2012)\citenamefont
  {Lherbier}, \citenamefont {Dubois}, \citenamefont {Declerck}, \citenamefont
  {Niquet}, \citenamefont {Roche},\ and\ \citenamefont
  {Charlier}}]{lherbier2012transport}%
  \BibitemOpen
  \bibfield  {author} {\bibinfo {author} {\bibfnamefont {A.}~\bibnamefont
  {Lherbier}}, \bibinfo {author} {\bibfnamefont {S.~M.-M.}\ \bibnamefont
  {Dubois}}, \bibinfo {author} {\bibfnamefont {X.}~\bibnamefont {Declerck}},
  \bibinfo {author} {\bibfnamefont {Y.-M.}\ \bibnamefont {Niquet}}, \bibinfo
  {author} {\bibfnamefont {S.}~\bibnamefont {Roche}},\ and\ \bibinfo {author}
  {\bibfnamefont {J.-C.}\ \bibnamefont {Charlier}},\ }\href@noop {} {\bibfield
  {journal} {\bibinfo  {journal} {Physical Review B}\ }\textbf {\bibinfo
  {volume} {86}},\ \bibinfo {pages} {075402} (\bibinfo {year}
  {2012})}\BibitemShut {NoStop}%
\bibitem [{\citenamefont {Si}\ \emph {et~al.}(2016)\citenamefont {Si},
  \citenamefont {Sun},\ and\ \citenamefont {Liu}}]{si2016strain}%
  \BibitemOpen
  \bibfield  {author} {\bibinfo {author} {\bibfnamefont {C.}~\bibnamefont
  {Si}}, \bibinfo {author} {\bibfnamefont {Z.}~\bibnamefont {Sun}},\ and\
  \bibinfo {author} {\bibfnamefont {F.}~\bibnamefont {Liu}},\ }\href@noop {}
  {\bibfield  {journal} {\bibinfo  {journal} {Nanoscale}\ }\textbf {\bibinfo
  {volume} {8}},\ \bibinfo {pages} {3207} (\bibinfo {year} {2016})}\BibitemShut
  {NoStop}%
\bibitem [{\citenamefont {Guinea}(2012)}]{guinea2012strain}%
  \BibitemOpen
  \bibfield  {author} {\bibinfo {author} {\bibfnamefont {F.}~\bibnamefont
  {Guinea}},\ }\href@noop {} {\bibfield  {journal} {\bibinfo  {journal} {Solid
  State Communications}\ }\textbf {\bibinfo {volume} {152}},\ \bibinfo {pages}
  {1437} (\bibinfo {year} {2012})}\BibitemShut {NoStop}%
\bibitem [{\citenamefont {Algara-Siller}\ \emph {et~al.}(2014)\citenamefont
  {Algara-Siller}, \citenamefont {Severin}, \citenamefont {Chong},
  \citenamefont {Bj{\"o}rkman}, \citenamefont {Palgrave}, \citenamefont
  {Laybourn}, \citenamefont {Antonietti}, \citenamefont {Khimyak},
  \citenamefont {Krasheninnikov}, \citenamefont {Rabe} \emph
  {et~al.}}]{algara2014triazine}%
  \BibitemOpen
  \bibfield  {author} {\bibinfo {author} {\bibfnamefont {G.}~\bibnamefont
  {Algara-Siller}}, \bibinfo {author} {\bibfnamefont {N.}~\bibnamefont
  {Severin}}, \bibinfo {author} {\bibfnamefont {S.~Y.}\ \bibnamefont {Chong}},
  \bibinfo {author} {\bibfnamefont {T.}~\bibnamefont {Bj{\"o}rkman}}, \bibinfo
  {author} {\bibfnamefont {R.~G.}\ \bibnamefont {Palgrave}}, \bibinfo {author}
  {\bibfnamefont {A.}~\bibnamefont {Laybourn}}, \bibinfo {author}
  {\bibfnamefont {M.}~\bibnamefont {Antonietti}}, \bibinfo {author}
  {\bibfnamefont {Y.~Z.}\ \bibnamefont {Khimyak}}, \bibinfo {author}
  {\bibfnamefont {A.~V.}\ \bibnamefont {Krasheninnikov}}, \bibinfo {author}
  {\bibfnamefont {J.~P.}\ \bibnamefont {Rabe}}, \emph {et~al.},\ }\href@noop {}
  {\bibfield  {journal} {\bibinfo  {journal} {Angewandte Chemie International
  Edition}\ }\textbf {\bibinfo {volume} {53}},\ \bibinfo {pages} {7450}
  (\bibinfo {year} {2014})}\BibitemShut {NoStop}%
\bibitem [{\citenamefont {Geim}\ and\ \citenamefont
  {Grigorieva}(2013)}]{geim2013van}%
  \BibitemOpen
  \bibfield  {author} {\bibinfo {author} {\bibfnamefont {A.~K.}\ \bibnamefont
  {Geim}}\ and\ \bibinfo {author} {\bibfnamefont {I.~V.}\ \bibnamefont
  {Grigorieva}},\ }\href@noop {} {\bibfield  {journal} {\bibinfo  {journal}
  {Nature}\ }\textbf {\bibinfo {volume} {499}},\ \bibinfo {pages} {419}
  (\bibinfo {year} {2013})}\BibitemShut {NoStop}%
\bibitem [{\citenamefont {Wang}\ \emph {et~al.}(2012)\citenamefont {Wang},
  \citenamefont {Kalantar-Zadeh}, \citenamefont {Kis}, \citenamefont
  {Coleman},\ and\ \citenamefont {Strano}}]{wang2012electronics}%
  \BibitemOpen
  \bibfield  {author} {\bibinfo {author} {\bibfnamefont {Q.~H.}\ \bibnamefont
  {Wang}}, \bibinfo {author} {\bibfnamefont {K.}~\bibnamefont
  {Kalantar-Zadeh}}, \bibinfo {author} {\bibfnamefont {A.}~\bibnamefont {Kis}},
  \bibinfo {author} {\bibfnamefont {J.~N.}\ \bibnamefont {Coleman}},\ and\
  \bibinfo {author} {\bibfnamefont {M.~S.}\ \bibnamefont {Strano}},\
  }\href@noop {} {\bibfield  {journal} {\bibinfo  {journal} {Nature
  nanotechnology}\ }\textbf {\bibinfo {volume} {7}},\ \bibinfo {pages} {699}
  (\bibinfo {year} {2012})}\BibitemShut {NoStop}%
\bibitem [{\citenamefont {Radisavljevic}\ \emph {et~al.}(2011)\citenamefont
  {Radisavljevic}, \citenamefont {Radenovic}, \citenamefont {Brivio},
  \citenamefont {Giacometti},\ and\ \citenamefont
  {Kis}}]{radisavljevic2011single}%
  \BibitemOpen
  \bibfield  {author} {\bibinfo {author} {\bibfnamefont {B.}~\bibnamefont
  {Radisavljevic}}, \bibinfo {author} {\bibfnamefont {A.}~\bibnamefont
  {Radenovic}}, \bibinfo {author} {\bibfnamefont {J.}~\bibnamefont {Brivio}},
  \bibinfo {author} {\bibfnamefont {V.}~\bibnamefont {Giacometti}},\ and\
  \bibinfo {author} {\bibfnamefont {A.}~\bibnamefont {Kis}},\ }\href@noop {}
  {\bibfield  {journal} {\bibinfo  {journal} {Nature nanotechnology}\ }\textbf
  {\bibinfo {volume} {6}},\ \bibinfo {pages} {147} (\bibinfo {year}
  {2011})}\BibitemShut {NoStop}%
\bibitem [{\citenamefont {Das}\ \emph {et~al.}(2014)\citenamefont {Das},
  \citenamefont {Demarteau},\ and\ \citenamefont {Roelofs}}]{das2014ambipolar}%
  \BibitemOpen
  \bibfield  {author} {\bibinfo {author} {\bibfnamefont {S.}~\bibnamefont
  {Das}}, \bibinfo {author} {\bibfnamefont {M.}~\bibnamefont {Demarteau}},\
  and\ \bibinfo {author} {\bibfnamefont {A.}~\bibnamefont {Roelofs}},\
  }\href@noop {} {\bibfield  {journal} {\bibinfo  {journal} {ACS nano}\
  }\textbf {\bibinfo {volume} {8}},\ \bibinfo {pages} {11730} (\bibinfo {year}
  {2014})}\BibitemShut {NoStop}%
\bibitem [{\citenamefont {Li}\ \emph {et~al.}(2014{\natexlab{a}})\citenamefont
  {Li}, \citenamefont {Yu}, \citenamefont {Ye}, \citenamefont {Ge},
  \citenamefont {Ou}, \citenamefont {Wu}, \citenamefont {Feng}, \citenamefont
  {Chen},\ and\ \citenamefont {Zhang}}]{li2014black}%
  \BibitemOpen
  \bibfield  {author} {\bibinfo {author} {\bibfnamefont {L.}~\bibnamefont
  {Li}}, \bibinfo {author} {\bibfnamefont {Y.}~\bibnamefont {Yu}}, \bibinfo
  {author} {\bibfnamefont {G.~J.}\ \bibnamefont {Ye}}, \bibinfo {author}
  {\bibfnamefont {Q.}~\bibnamefont {Ge}}, \bibinfo {author} {\bibfnamefont
  {X.}~\bibnamefont {Ou}}, \bibinfo {author} {\bibfnamefont {H.}~\bibnamefont
  {Wu}}, \bibinfo {author} {\bibfnamefont {D.}~\bibnamefont {Feng}}, \bibinfo
  {author} {\bibfnamefont {X.~H.}\ \bibnamefont {Chen}},\ and\ \bibinfo
  {author} {\bibfnamefont {Y.}~\bibnamefont {Zhang}},\ }\href@noop {}
  {\bibfield  {journal} {\bibinfo  {journal} {Nature nanotechnology}\ }\textbf
  {\bibinfo {volume} {9}},\ \bibinfo {pages} {372} (\bibinfo {year}
  {2014}{\natexlab{a}})}\BibitemShut {NoStop}%
\bibitem [{\citenamefont {Bandurin}\ \emph {et~al.}(2017)\citenamefont
  {Bandurin}, \citenamefont {Tyurnina}, \citenamefont {Geliang}, \citenamefont
  {Mishchenko}, \citenamefont {Z{\'o}lyomi}, \citenamefont {Morozov},
  \citenamefont {Kumar}, \citenamefont {Gorbachev}, \citenamefont {Kudrynskyi},
  \citenamefont {Pezzini} \emph {et~al.}}]{bandurin2017high}%
  \BibitemOpen
  \bibfield  {author} {\bibinfo {author} {\bibfnamefont {D.~A.}\ \bibnamefont
  {Bandurin}}, \bibinfo {author} {\bibfnamefont {A.~V.}\ \bibnamefont
  {Tyurnina}}, \bibinfo {author} {\bibfnamefont {L.~Y.}\ \bibnamefont
  {Geliang}}, \bibinfo {author} {\bibfnamefont {A.}~\bibnamefont {Mishchenko}},
  \bibinfo {author} {\bibfnamefont {V.}~\bibnamefont {Z{\'o}lyomi}}, \bibinfo
  {author} {\bibfnamefont {S.~V.}\ \bibnamefont {Morozov}}, \bibinfo {author}
  {\bibfnamefont {R.~K.}\ \bibnamefont {Kumar}}, \bibinfo {author}
  {\bibfnamefont {R.~V.}\ \bibnamefont {Gorbachev}}, \bibinfo {author}
  {\bibfnamefont {Z.~R.}\ \bibnamefont {Kudrynskyi}}, \bibinfo {author}
  {\bibfnamefont {S.}~\bibnamefont {Pezzini}}, \emph {et~al.},\ }\href@noop {}
  {\bibfield  {journal} {\bibinfo  {journal} {Nature nanotechnology}\ }\textbf
  {\bibinfo {volume} {12}},\ \bibinfo {pages} {223} (\bibinfo {year}
  {2017})}\BibitemShut {NoStop}%
\bibitem [{\citenamefont {Hong}\ \emph {et~al.}(2020)\citenamefont {Hong},
  \citenamefont {Liu}, \citenamefont {Wang}, \citenamefont {Zhou},
  \citenamefont {Ma}, \citenamefont {Xu}, \citenamefont {Feng}, \citenamefont
  {Chen}, \citenamefont {Chen}, \citenamefont {Sun} \emph
  {et~al.}}]{hong2020chemical}%
  \BibitemOpen
  \bibfield  {author} {\bibinfo {author} {\bibfnamefont {Y.-L.}\ \bibnamefont
  {Hong}}, \bibinfo {author} {\bibfnamefont {Z.}~\bibnamefont {Liu}}, \bibinfo
  {author} {\bibfnamefont {L.}~\bibnamefont {Wang}}, \bibinfo {author}
  {\bibfnamefont {T.}~\bibnamefont {Zhou}}, \bibinfo {author} {\bibfnamefont
  {W.}~\bibnamefont {Ma}}, \bibinfo {author} {\bibfnamefont {C.}~\bibnamefont
  {Xu}}, \bibinfo {author} {\bibfnamefont {S.}~\bibnamefont {Feng}}, \bibinfo
  {author} {\bibfnamefont {L.}~\bibnamefont {Chen}}, \bibinfo {author}
  {\bibfnamefont {M.-L.}\ \bibnamefont {Chen}}, \bibinfo {author}
  {\bibfnamefont {D.-M.}\ \bibnamefont {Sun}}, \emph {et~al.},\ }\href@noop {}
  {\bibfield  {journal} {\bibinfo  {journal} {Science}\ }\textbf {\bibinfo
  {volume} {369}},\ \bibinfo {pages} {670} (\bibinfo {year}
  {2020})}\BibitemShut {NoStop}%
\bibitem [{\citenamefont {Mahmood}\ \emph {et~al.}(2016)\citenamefont
  {Mahmood}, \citenamefont {Lee}, \citenamefont {Jung}, \citenamefont {Shin},
  \citenamefont {Choi}, \citenamefont {Seo}, \citenamefont {Jung},
  \citenamefont {Kim}, \citenamefont {Li}, \citenamefont {Lah} \emph
  {et~al.}}]{mahmood2016two}%
  \BibitemOpen
  \bibfield  {author} {\bibinfo {author} {\bibfnamefont {J.}~\bibnamefont
  {Mahmood}}, \bibinfo {author} {\bibfnamefont {E.~K.}\ \bibnamefont {Lee}},
  \bibinfo {author} {\bibfnamefont {M.}~\bibnamefont {Jung}}, \bibinfo {author}
  {\bibfnamefont {D.}~\bibnamefont {Shin}}, \bibinfo {author} {\bibfnamefont
  {H.-J.}\ \bibnamefont {Choi}}, \bibinfo {author} {\bibfnamefont {J.-M.}\
  \bibnamefont {Seo}}, \bibinfo {author} {\bibfnamefont {S.-M.}\ \bibnamefont
  {Jung}}, \bibinfo {author} {\bibfnamefont {D.}~\bibnamefont {Kim}}, \bibinfo
  {author} {\bibfnamefont {F.}~\bibnamefont {Li}}, \bibinfo {author}
  {\bibfnamefont {M.~S.}\ \bibnamefont {Lah}}, \emph {et~al.},\ }\href@noop {}
  {\bibfield  {journal} {\bibinfo  {journal} {Proceedings of the National
  Academy of Sciences}\ }\textbf {\bibinfo {volume} {113}},\ \bibinfo {pages}
  {7414} (\bibinfo {year} {2016})}\BibitemShut {NoStop}%
\bibitem [{\citenamefont {Seo}\ \emph {et~al.}(2021)\citenamefont {Seo},
  \citenamefont {Lee}, \citenamefont {Lee}, \citenamefont {Hwang},
  \citenamefont {Son}, \citenamefont {Ahn}, \citenamefont {Cho},\ and\
  \citenamefont {Kim}}]{seo2021dominant}%
  \BibitemOpen
  \bibfield  {author} {\bibinfo {author} {\bibfnamefont {T.~H.}\ \bibnamefont
  {Seo}}, \bibinfo {author} {\bibfnamefont {W.}~\bibnamefont {Lee}}, \bibinfo
  {author} {\bibfnamefont {K.~S.}\ \bibnamefont {Lee}}, \bibinfo {author}
  {\bibfnamefont {J.~Y.}\ \bibnamefont {Hwang}}, \bibinfo {author}
  {\bibfnamefont {D.~I.}\ \bibnamefont {Son}}, \bibinfo {author} {\bibfnamefont
  {S.}~\bibnamefont {Ahn}}, \bibinfo {author} {\bibfnamefont {H.}~\bibnamefont
  {Cho}},\ and\ \bibinfo {author} {\bibfnamefont {M.~J.}\ \bibnamefont {Kim}},\
  }\href@noop {} {\bibfield  {journal} {\bibinfo  {journal} {Carbon}\ }\textbf
  {\bibinfo {volume} {182}},\ \bibinfo {pages} {791} (\bibinfo {year}
  {2021})}\BibitemShut {NoStop}%
\bibitem [{\citenamefont {Bykov}\ \emph {et~al.}(2021)\citenamefont {Bykov},
  \citenamefont {Bykova}, \citenamefont {Ponomareva}, \citenamefont {Tasnadi},
  \citenamefont {Chariton}, \citenamefont {Prakapenka}, \citenamefont
  {Glazyrin}, \citenamefont {Smith}, \citenamefont {Mahmood}, \citenamefont
  {Abrikosov} \emph {et~al.}}]{bykov2021realization}%
  \BibitemOpen
  \bibfield  {author} {\bibinfo {author} {\bibfnamefont {M.}~\bibnamefont
  {Bykov}}, \bibinfo {author} {\bibfnamefont {E.}~\bibnamefont {Bykova}},
  \bibinfo {author} {\bibfnamefont {A.~V.}\ \bibnamefont {Ponomareva}},
  \bibinfo {author} {\bibfnamefont {F.}~\bibnamefont {Tasnadi}}, \bibinfo
  {author} {\bibfnamefont {S.}~\bibnamefont {Chariton}}, \bibinfo {author}
  {\bibfnamefont {V.~B.}\ \bibnamefont {Prakapenka}}, \bibinfo {author}
  {\bibfnamefont {K.}~\bibnamefont {Glazyrin}}, \bibinfo {author}
  {\bibfnamefont {J.~S.}\ \bibnamefont {Smith}}, \bibinfo {author}
  {\bibfnamefont {M.~F.}\ \bibnamefont {Mahmood}}, \bibinfo {author}
  {\bibfnamefont {I.~A.}\ \bibnamefont {Abrikosov}}, \emph {et~al.},\
  }\href@noop {} {\bibfield  {journal} {\bibinfo  {journal} {ACS nano}\
  }\textbf {\bibinfo {volume} {15}},\ \bibinfo {pages} {13539} (\bibinfo {year}
  {2021})}\BibitemShut {NoStop}%
\bibitem [{\citenamefont {Fang}\ \emph {et~al.}(2021)\citenamefont {Fang},
  \citenamefont {Wang}, \citenamefont {Wang}, \citenamefont {Zhai},\ and\
  \citenamefont {Huang}}]{fang20212d}%
  \BibitemOpen
  \bibfield  {author} {\bibinfo {author} {\bibfnamefont {Y.}~\bibnamefont
  {Fang}}, \bibinfo {author} {\bibfnamefont {F.}~\bibnamefont {Wang}}, \bibinfo
  {author} {\bibfnamefont {R.}~\bibnamefont {Wang}}, \bibinfo {author}
  {\bibfnamefont {T.}~\bibnamefont {Zhai}},\ and\ \bibinfo {author}
  {\bibfnamefont {F.}~\bibnamefont {Huang}},\ }\href@noop {} {\bibfield
  {journal} {\bibinfo  {journal} {Advanced Materials}\ ,\ \bibinfo {pages}
  {2101505}} (\bibinfo {year} {2021})}\BibitemShut {NoStop}%
\bibitem [{\citenamefont {Wang}\ \emph {et~al.}(2022)\citenamefont {Wang},
  \citenamefont {Xiong}, \citenamefont {Zhao}, \citenamefont {Zhou},
  \citenamefont {Xin}, \citenamefont {Deng}, \citenamefont {Kang},
  \citenamefont {Yang}, \citenamefont {Liu},\ and\ \citenamefont
  {Wei}}]{wang2022polarimetric}%
  \BibitemOpen
  \bibfield  {author} {\bibinfo {author} {\bibfnamefont {X.}~\bibnamefont
  {Wang}}, \bibinfo {author} {\bibfnamefont {T.}~\bibnamefont {Xiong}},
  \bibinfo {author} {\bibfnamefont {K.}~\bibnamefont {Zhao}}, \bibinfo {author}
  {\bibfnamefont {Z.}~\bibnamefont {Zhou}}, \bibinfo {author} {\bibfnamefont
  {K.}~\bibnamefont {Xin}}, \bibinfo {author} {\bibfnamefont {H.-X.}\
  \bibnamefont {Deng}}, \bibinfo {author} {\bibfnamefont {J.}~\bibnamefont
  {Kang}}, \bibinfo {author} {\bibfnamefont {J.}~\bibnamefont {Yang}}, \bibinfo
  {author} {\bibfnamefont {Y.-Y.}\ \bibnamefont {Liu}},\ and\ \bibinfo {author}
  {\bibfnamefont {Z.}~\bibnamefont {Wei}},\ }\href@noop {} {\bibfield
  {journal} {\bibinfo  {journal} {Advanced Materials}\ }\textbf {\bibinfo
  {volume} {34}},\ \bibinfo {pages} {2107206} (\bibinfo {year}
  {2022})}\BibitemShut {NoStop}%
\bibitem [{\citenamefont {Li}\ \emph {et~al.}(2021)\citenamefont {Li},
  \citenamefont {Zhang}, \citenamefont {Zhu}, \citenamefont {Shen},
  \citenamefont {Hu}, \citenamefont {Fu}, \citenamefont {Yan}, \citenamefont
  {Zhou}, \citenamefont {Zheng}, \citenamefont {Lei} \emph
  {et~al.}}]{li2021penta}%
  \BibitemOpen
  \bibfield  {author} {\bibinfo {author} {\bibfnamefont {P.}~\bibnamefont
  {Li}}, \bibinfo {author} {\bibfnamefont {J.}~\bibnamefont {Zhang}}, \bibinfo
  {author} {\bibfnamefont {C.}~\bibnamefont {Zhu}}, \bibinfo {author}
  {\bibfnamefont {W.}~\bibnamefont {Shen}}, \bibinfo {author} {\bibfnamefont
  {C.}~\bibnamefont {Hu}}, \bibinfo {author} {\bibfnamefont {W.}~\bibnamefont
  {Fu}}, \bibinfo {author} {\bibfnamefont {L.}~\bibnamefont {Yan}}, \bibinfo
  {author} {\bibfnamefont {L.}~\bibnamefont {Zhou}}, \bibinfo {author}
  {\bibfnamefont {L.}~\bibnamefont {Zheng}}, \bibinfo {author} {\bibfnamefont
  {H.}~\bibnamefont {Lei}}, \emph {et~al.},\ }\href@noop {} {\bibfield
  {journal} {\bibinfo  {journal} {Advanced Materials}\ }\textbf {\bibinfo
  {volume} {33}},\ \bibinfo {pages} {2102541} (\bibinfo {year}
  {2021})}\BibitemShut {NoStop}%
\bibitem [{\citenamefont {Zhang}\ \emph {et~al.}(2021)\citenamefont {Zhang},
  \citenamefont {Zhang}, \citenamefont {Si}, \citenamefont {Li}, \citenamefont
  {Deng}, \citenamefont {Yang}, \citenamefont {Liu}, \citenamefont {Dai},\ and\
  \citenamefont {Luo}}]{Zhang2021ACS}%
  \BibitemOpen
  \bibfield  {author} {\bibinfo {author} {\bibfnamefont {S.}~\bibnamefont
  {Zhang}}, \bibinfo {author} {\bibfnamefont {Z.}~\bibnamefont {Zhang}},
  \bibinfo {author} {\bibfnamefont {Y.}~\bibnamefont {Si}}, \bibinfo {author}
  {\bibfnamefont {B.}~\bibnamefont {Li}}, \bibinfo {author} {\bibfnamefont
  {F.}~\bibnamefont {Deng}}, \bibinfo {author} {\bibfnamefont {L.}~\bibnamefont
  {Yang}}, \bibinfo {author} {\bibfnamefont {X.}~\bibnamefont {Liu}}, \bibinfo
  {author} {\bibfnamefont {W.}~\bibnamefont {Dai}},\ and\ \bibinfo {author}
  {\bibfnamefont {S.~a.}\ \bibnamefont {Luo}},\ }\href@noop {} {\bibfield
  {journal} {\bibinfo  {journal} {ACS Nano}\ }\textbf {\bibinfo {volume}
  {15}},\ \bibinfo {pages} {15238} (\bibinfo {year} {2021})}\BibitemShut
  {NoStop}%
\bibitem [{\citenamefont {Kresse}\ and\ \citenamefont
  {Furthm{\"u}ller}(1996)}]{kresse1996efficient}%
  \BibitemOpen
  \bibfield  {author} {\bibinfo {author} {\bibfnamefont {G.}~\bibnamefont
  {Kresse}}\ and\ \bibinfo {author} {\bibfnamefont {J.}~\bibnamefont
  {Furthm{\"u}ller}},\ }\href@noop {} {\bibfield  {journal} {\bibinfo
  {journal} {Physical review B}\ }\textbf {\bibinfo {volume} {54}},\ \bibinfo
  {pages} {11169} (\bibinfo {year} {1996})}\BibitemShut {NoStop}%
\bibitem [{\citenamefont {Perdew}\ \emph {et~al.}(1996)\citenamefont {Perdew},
  \citenamefont {Burke},\ and\ \citenamefont
  {Ernzerhof}}]{perdew1996generalized}%
  \BibitemOpen
  \bibfield  {author} {\bibinfo {author} {\bibfnamefont {J.~P.}\ \bibnamefont
  {Perdew}}, \bibinfo {author} {\bibfnamefont {K.}~\bibnamefont {Burke}},\ and\
  \bibinfo {author} {\bibfnamefont {M.}~\bibnamefont {Ernzerhof}},\ }\href@noop
  {} {\bibfield  {journal} {\bibinfo  {journal} {Physical review letters}\
  }\textbf {\bibinfo {volume} {77}},\ \bibinfo {pages} {3865} (\bibinfo {year}
  {1996})}\BibitemShut {NoStop}%
\bibitem [{\citenamefont {Monkhorst}\ and\ \citenamefont
  {Pack}(1976)}]{monkhorst1976special}%
  \BibitemOpen
  \bibfield  {author} {\bibinfo {author} {\bibfnamefont {H.~J.}\ \bibnamefont
  {Monkhorst}}\ and\ \bibinfo {author} {\bibfnamefont {J.~D.}\ \bibnamefont
  {Pack}},\ }\href@noop {} {\bibfield  {journal} {\bibinfo  {journal} {Physical
  review B}\ }\textbf {\bibinfo {volume} {13}},\ \bibinfo {pages} {5188}
  (\bibinfo {year} {1976})}\BibitemShut {NoStop}%
\bibitem [{\citenamefont {Grimme}\ \emph {et~al.}(2010)\citenamefont {Grimme},
  \citenamefont {Antony}, \citenamefont {Ehrlich},\ and\ \citenamefont
  {Krieg}}]{grimme2010consistent}%
  \BibitemOpen
  \bibfield  {author} {\bibinfo {author} {\bibfnamefont {S.}~\bibnamefont
  {Grimme}}, \bibinfo {author} {\bibfnamefont {J.}~\bibnamefont {Antony}},
  \bibinfo {author} {\bibfnamefont {S.}~\bibnamefont {Ehrlich}},\ and\ \bibinfo
  {author} {\bibfnamefont {H.}~\bibnamefont {Krieg}},\ }\href@noop {}
  {\bibfield  {journal} {\bibinfo  {journal} {The Journal of chemical physics}\
  }\textbf {\bibinfo {volume} {132}},\ \bibinfo {pages} {154104} (\bibinfo
  {year} {2010})}\BibitemShut {NoStop}%
\bibitem [{\citenamefont {Krukau}\ \emph {et~al.}(2006)\citenamefont {Krukau},
  \citenamefont {Vydrov}, \citenamefont {Izmaylov},\ and\ \citenamefont
  {Scuseria}}]{krukau2006influence}%
  \BibitemOpen
  \bibfield  {author} {\bibinfo {author} {\bibfnamefont {A.~V.}\ \bibnamefont
  {Krukau}}, \bibinfo {author} {\bibfnamefont {O.~A.}\ \bibnamefont {Vydrov}},
  \bibinfo {author} {\bibfnamefont {A.~F.}\ \bibnamefont {Izmaylov}},\ and\
  \bibinfo {author} {\bibfnamefont {G.~E.}\ \bibnamefont {Scuseria}},\
  }\href@noop {} {\bibfield  {journal} {\bibinfo  {journal} {The Journal of
  chemical physics}\ }\textbf {\bibinfo {volume} {125}},\ \bibinfo {pages}
  {224106} (\bibinfo {year} {2006})}\BibitemShut {NoStop}%
\bibitem [{\citenamefont {Scheidemantel}\ \emph {et~al.}(2003)\citenamefont
  {Scheidemantel}, \citenamefont {Ambrosch-Draxl}, \citenamefont {Thonhauser},
  \citenamefont {Badding},\ and\ \citenamefont {Sofo}}]{scheidee}%
  \BibitemOpen
  \bibfield  {author} {\bibinfo {author} {\bibfnamefont {T.}~\bibnamefont
  {Scheidemantel}}, \bibinfo {author} {\bibfnamefont {C.}~\bibnamefont
  {Ambrosch-Draxl}}, \bibinfo {author} {\bibfnamefont {T.}~\bibnamefont
  {Thonhauser}}, \bibinfo {author} {\bibfnamefont {J.}~\bibnamefont
  {Badding}},\ and\ \bibinfo {author} {\bibfnamefont {J.~O.}\ \bibnamefont
  {Sofo}},\ }\href@noop {} {\bibfield  {journal} {\bibinfo  {journal} {Physical
  Review B}\ }\textbf {\bibinfo {volume} {68}},\ \bibinfo {pages} {125210}
  (\bibinfo {year} {2003})}\BibitemShut {NoStop}%
\bibitem [{\citenamefont {Madsen}\ and\ \citenamefont
  {Singh}(2006)}]{madsen2006bol}%
  \BibitemOpen
  \bibfield  {author} {\bibinfo {author} {\bibfnamefont {G.~K.}\ \bibnamefont
  {Madsen}}\ and\ \bibinfo {author} {\bibfnamefont {D.~J.}\ \bibnamefont
  {Singh}},\ }\href@noop {} {\bibfield  {journal} {\bibinfo  {journal}
  {Computer Physics Communications}\ }\textbf {\bibinfo {volume} {175}},\
  \bibinfo {pages} {67} (\bibinfo {year} {2006})}\BibitemShut {NoStop}%
\bibitem [{\citenamefont {Togo}\ and\ \citenamefont
  {Tanaka}(2015)}]{togo2015first}%
  \BibitemOpen
  \bibfield  {author} {\bibinfo {author} {\bibfnamefont {A.}~\bibnamefont
  {Togo}}\ and\ \bibinfo {author} {\bibfnamefont {I.}~\bibnamefont {Tanaka}},\
  }\href@noop {} {\bibfield  {journal} {\bibinfo  {journal} {Scripta
  Materialia}\ }\textbf {\bibinfo {volume} {108}},\ \bibinfo {pages} {1}
  (\bibinfo {year} {2015})}\BibitemShut {NoStop}%
\bibitem [{\citenamefont {Shapeev}(2016)}]{shapeev2016moment}%
  \BibitemOpen
  \bibfield  {author} {\bibinfo {author} {\bibfnamefont {A.~V.}\ \bibnamefont
  {Shapeev}},\ }\href@noop {} {\bibfield  {journal} {\bibinfo  {journal}
  {Multiscale Modeling \& Simulation}\ }\textbf {\bibinfo {volume} {14}},\
  \bibinfo {pages} {1153} (\bibinfo {year} {2016})}\BibitemShut {NoStop}%
\bibitem [{\citenamefont {Mortazavi}\ \emph {et~al.}(2020)\citenamefont
  {Mortazavi}, \citenamefont {Novikov}, \citenamefont {Podryabinkin},
  \citenamefont {Roche}, \citenamefont {Rabczuk}, \citenamefont {Shapeev},\
  and\ \citenamefont {Zhuang}}]{mortazavi2020exploring}%
  \BibitemOpen
  \bibfield  {author} {\bibinfo {author} {\bibfnamefont {B.}~\bibnamefont
  {Mortazavi}}, \bibinfo {author} {\bibfnamefont {I.~S.}\ \bibnamefont
  {Novikov}}, \bibinfo {author} {\bibfnamefont {E.~V.}\ \bibnamefont
  {Podryabinkin}}, \bibinfo {author} {\bibfnamefont {S.}~\bibnamefont {Roche}},
  \bibinfo {author} {\bibfnamefont {T.}~\bibnamefont {Rabczuk}}, \bibinfo
  {author} {\bibfnamefont {A.~V.}\ \bibnamefont {Shapeev}},\ and\ \bibinfo
  {author} {\bibfnamefont {X.}~\bibnamefont {Zhuang}},\ }\href@noop {}
  {\bibfield  {journal} {\bibinfo  {journal} {Applied Materials Today}\
  }\textbf {\bibinfo {volume} {20}},\ \bibinfo {pages} {100685} (\bibinfo
  {year} {2020})}\BibitemShut {NoStop}%
\bibitem [{\citenamefont {Novikov}\ \emph {et~al.}(2020)\citenamefont
  {Novikov}, \citenamefont {Gubaev}, \citenamefont {Podryabinkin},\ and\
  \citenamefont {Shapeev}}]{novikov2020mlip}%
  \BibitemOpen
  \bibfield  {author} {\bibinfo {author} {\bibfnamefont {I.~S.}\ \bibnamefont
  {Novikov}}, \bibinfo {author} {\bibfnamefont {K.}~\bibnamefont {Gubaev}},
  \bibinfo {author} {\bibfnamefont {E.~V.}\ \bibnamefont {Podryabinkin}},\ and\
  \bibinfo {author} {\bibfnamefont {A.~V.}\ \bibnamefont {Shapeev}},\
  }\href@noop {} {\bibfield  {journal} {\bibinfo  {journal} {Machine Learning:
  Science and Technology}\ }\textbf {\bibinfo {volume} {2}},\ \bibinfo {pages}
  {025002} (\bibinfo {year} {2020})}\BibitemShut {NoStop}%
\bibitem [{\citenamefont {Li}\ \emph {et~al.}(2014{\natexlab{b}})\citenamefont
  {Li}, \citenamefont {Carrete}, \citenamefont {Katcho},\ and\ \citenamefont
  {Mingo}}]{li2014shengbte}%
  \BibitemOpen
  \bibfield  {author} {\bibinfo {author} {\bibfnamefont {W.}~\bibnamefont
  {Li}}, \bibinfo {author} {\bibfnamefont {J.}~\bibnamefont {Carrete}},
  \bibinfo {author} {\bibfnamefont {N.~A.}\ \bibnamefont {Katcho}},\ and\
  \bibinfo {author} {\bibfnamefont {N.}~\bibnamefont {Mingo}},\ }\href@noop {}
  {\bibfield  {journal} {\bibinfo  {journal} {Computer Physics Communications}\
  }\textbf {\bibinfo {volume} {185}},\ \bibinfo {pages} {1747} (\bibinfo {year}
  {2014}{\natexlab{b}})}\BibitemShut {NoStop}%
\bibitem [{\citenamefont {Mortazavi}\ \emph {et~al.}(2021)\citenamefont
  {Mortazavi}, \citenamefont {Podryabinkin}, \citenamefont {Novikov},
  \citenamefont {Rabczuk}, \citenamefont {Zhuang},\ and\ \citenamefont
  {Shapeev}}]{mortazavi2021accelerating}%
  \BibitemOpen
  \bibfield  {author} {\bibinfo {author} {\bibfnamefont {B.}~\bibnamefont
  {Mortazavi}}, \bibinfo {author} {\bibfnamefont {E.~V.}\ \bibnamefont
  {Podryabinkin}}, \bibinfo {author} {\bibfnamefont {I.~S.}\ \bibnamefont
  {Novikov}}, \bibinfo {author} {\bibfnamefont {T.}~\bibnamefont {Rabczuk}},
  \bibinfo {author} {\bibfnamefont {X.}~\bibnamefont {Zhuang}},\ and\ \bibinfo
  {author} {\bibfnamefont {A.~V.}\ \bibnamefont {Shapeev}},\ }\href@noop {}
  {\bibfield  {journal} {\bibinfo  {journal} {Computer Physics Communications}\
  }\textbf {\bibinfo {volume} {258}},\ \bibinfo {pages} {107583} (\bibinfo
  {year} {2021})}\BibitemShut {NoStop}%
\bibitem [{\citenamefont {Mohebpour}\ \emph
  {et~al.}(2020{\natexlab{a}})\citenamefont {Mohebpour}, \citenamefont
  {Vishkayi},\ and\ \citenamefont {Tagani}}]{mohebpour2020tuning}%
  \BibitemOpen
  \bibfield  {author} {\bibinfo {author} {\bibfnamefont {M.~A.}\ \bibnamefont
  {Mohebpour}}, \bibinfo {author} {\bibfnamefont {S.~I.}\ \bibnamefont
  {Vishkayi}},\ and\ \bibinfo {author} {\bibfnamefont {M.~B.}\ \bibnamefont
  {Tagani}},\ }\href@noop {} {\bibfield  {journal} {\bibinfo  {journal}
  {Journal of Applied Physics}\ }\textbf {\bibinfo {volume} {127}},\ \bibinfo
  {pages} {014302} (\bibinfo {year} {2020}{\natexlab{a}})}\BibitemShut
  {NoStop}%
\bibitem [{\citenamefont {Mohebpour}\ \emph {et~al.}(2022)\citenamefont
  {Mohebpour}, \citenamefont {Mozvashi}, \citenamefont {Vishkayi},\ and\
  \citenamefont {Tagani}}]{moheb2022tr}%
  \BibitemOpen
  \bibfield  {author} {\bibinfo {author} {\bibfnamefont {M.~A.}\ \bibnamefont
  {Mohebpour}}, \bibinfo {author} {\bibfnamefont {S.~M.}\ \bibnamefont
  {Mozvashi}}, \bibinfo {author} {\bibfnamefont {S.~I.}\ \bibnamefont
  {Vishkayi}},\ and\ \bibinfo {author} {\bibfnamefont {M.~B.}\ \bibnamefont
  {Tagani}},\ }\href@noop {} {\bibfield  {journal} {\bibinfo  {journal}
  {Physical Review Materials}\ }\textbf {\bibinfo {volume} {6}},\ \bibinfo
  {pages} {014012} (\bibinfo {year} {2022})}\BibitemShut {NoStop}%
\bibitem [{\citenamefont {Chang}\ \emph {et~al.}(2021)\citenamefont {Chang},
  \citenamefont {Wang}, \citenamefont {Song}, \citenamefont {Zhong},
  \citenamefont {Shi},\ and\ \citenamefont {Qian}}]{chang2021origin}%
  \BibitemOpen
  \bibfield  {author} {\bibinfo {author} {\bibfnamefont {H.}~\bibnamefont
  {Chang}}, \bibinfo {author} {\bibfnamefont {H.}~\bibnamefont {Wang}},
  \bibinfo {author} {\bibfnamefont {K.-K.}\ \bibnamefont {Song}}, \bibinfo
  {author} {\bibfnamefont {M.}~\bibnamefont {Zhong}}, \bibinfo {author}
  {\bibfnamefont {L.-B.}\ \bibnamefont {Shi}},\ and\ \bibinfo {author}
  {\bibfnamefont {P.}~\bibnamefont {Qian}},\ }\href@noop {} {\bibfield
  {journal} {\bibinfo  {journal} {Journal of Physics: Condensed Matter}\
  }\textbf {\bibinfo {volume} {34}},\ \bibinfo {pages} {013003} (\bibinfo
  {year} {2021})}\BibitemShut {NoStop}%
\bibitem [{\citenamefont {Su}\ \emph {et~al.}(2021)\citenamefont {Su},
  \citenamefont {Cao}, \citenamefont {Shi},\ and\ \citenamefont
  {Qian}}]{su2021investigation}%
  \BibitemOpen
  \bibfield  {author} {\bibinfo {author} {\bibfnamefont {Y.}~\bibnamefont
  {Su}}, \bibinfo {author} {\bibfnamefont {S.}~\bibnamefont {Cao}}, \bibinfo
  {author} {\bibfnamefont {L.-B.}\ \bibnamefont {Shi}},\ and\ \bibinfo {author}
  {\bibfnamefont {P.}~\bibnamefont {Qian}},\ }\href@noop {} {\bibfield
  {journal} {\bibinfo  {journal} {Journal of Applied Physics}\ }\textbf
  {\bibinfo {volume} {130}},\ \bibinfo {pages} {195703} (\bibinfo {year}
  {2021})}\BibitemShut {NoStop}%
\bibitem [{\citenamefont {Bardeen}\ and\ \citenamefont
  {Shockley}(1950)}]{bardeen1950deformation}%
  \BibitemOpen
  \bibfield  {author} {\bibinfo {author} {\bibfnamefont {J.}~\bibnamefont
  {Bardeen}}\ and\ \bibinfo {author} {\bibfnamefont {W.}~\bibnamefont
  {Shockley}},\ }\href@noop {} {\bibfield  {journal} {\bibinfo  {journal}
  {Physical review}\ }\textbf {\bibinfo {volume} {80}},\ \bibinfo {pages} {72}
  (\bibinfo {year} {1950})}\BibitemShut {NoStop}%
\bibitem [{\citenamefont {Nian}\ \emph {et~al.}(2021)\citenamefont {Nian},
  \citenamefont {Wang},\ and\ \citenamefont {Dong}}]{nian2021thermoelectric}%
  \BibitemOpen
  \bibfield  {author} {\bibinfo {author} {\bibfnamefont {T.}~\bibnamefont
  {Nian}}, \bibinfo {author} {\bibfnamefont {Z.}~\bibnamefont {Wang}},\ and\
  \bibinfo {author} {\bibfnamefont {B.}~\bibnamefont {Dong}},\ }\href@noop {}
  {\bibfield  {journal} {\bibinfo  {journal} {Applied Physics Letters}\
  }\textbf {\bibinfo {volume} {118}},\ \bibinfo {pages} {033103} (\bibinfo
  {year} {2021})}\BibitemShut {NoStop}%
\bibitem [{\citenamefont {Mohebpour}\ \emph
  {et~al.}(2020{\natexlab{b}})\citenamefont {Mohebpour}, \citenamefont
  {Vishkayi},\ and\ \citenamefont {Tagani}}]{mohebpour2020thermoelectric}%
  \BibitemOpen
  \bibfield  {author} {\bibinfo {author} {\bibfnamefont {M.~A.}\ \bibnamefont
  {Mohebpour}}, \bibinfo {author} {\bibfnamefont {S.~I.}\ \bibnamefont
  {Vishkayi}},\ and\ \bibinfo {author} {\bibfnamefont {M.~B.}\ \bibnamefont
  {Tagani}},\ }\href@noop {} {\bibfield  {journal} {\bibinfo  {journal}
  {Physical Chemistry Chemical Physics}\ }\textbf {\bibinfo {volume} {22}},\
  \bibinfo {pages} {23246} (\bibinfo {year} {2020}{\natexlab{b}})}\BibitemShut
  {NoStop}%
\bibitem [{\citenamefont {Naghavi}\ \emph {et~al.}(2018)\citenamefont
  {Naghavi}, \citenamefont {He}, \citenamefont {Xia},\ and\ \citenamefont
  {Wolverton}}]{naghavi2018pd2se3}%
  \BibitemOpen
  \bibfield  {author} {\bibinfo {author} {\bibfnamefont {S.~S.}\ \bibnamefont
  {Naghavi}}, \bibinfo {author} {\bibfnamefont {J.}~\bibnamefont {He}},
  \bibinfo {author} {\bibfnamefont {Y.}~\bibnamefont {Xia}},\ and\ \bibinfo
  {author} {\bibfnamefont {C.}~\bibnamefont {Wolverton}},\ }\href@noop {}
  {\bibfield  {journal} {\bibinfo  {journal} {Chemistry of Materials}\ }\textbf
  {\bibinfo {volume} {30}},\ \bibinfo {pages} {5639} (\bibinfo {year}
  {2018})}\BibitemShut {NoStop}%
\bibitem [{\citenamefont {Patel}\ \emph {et~al.}(2020)\citenamefont {Patel},
  \citenamefont {Singh}, \citenamefont {Sonvane}, \citenamefont {Thakor},\ and\
  \citenamefont {Ahuja}}]{patel2020high}%
  \BibitemOpen
  \bibfield  {author} {\bibinfo {author} {\bibfnamefont {A.}~\bibnamefont
  {Patel}}, \bibinfo {author} {\bibfnamefont {D.}~\bibnamefont {Singh}},
  \bibinfo {author} {\bibfnamefont {Y.}~\bibnamefont {Sonvane}}, \bibinfo
  {author} {\bibfnamefont {P.}~\bibnamefont {Thakor}},\ and\ \bibinfo {author}
  {\bibfnamefont {R.}~\bibnamefont {Ahuja}},\ }\href@noop {} {\bibfield
  {journal} {\bibinfo  {journal} {ACS applied materials \& interfaces}\
  }\textbf {\bibinfo {volume} {12}},\ \bibinfo {pages} {46212} (\bibinfo {year}
  {2020})}\BibitemShut {NoStop}%
\end{thebibliography}%
\end{document}